\documentclass[12pt]{article}
\usepackage{amsmath}
\usepackage{graphicx}
\usepackage{amssymb}
\usepackage{geometry}
\usepackage{epsfig}
\begin{document}
\begin{center}
\begin{large}
\textbf{Magnetisation reversal in Ising Ferromagnet by Thermal
and Field gradients}
\end{large}\end{center}
\vspace{1.5cm}
\begin{center}\emph{Ranajay Datta$^{1,2,*}$, Muktish Acharyya$^{2,+,a}$ and Abyaya Dhar$^{2,3,\dagger}$}\\
\vspace{.4cm} 
\emph{$^1$School of Physics, University of Hyderabad,\\
 Hyderabad-500046, INDIA}\\
\emph{$^2$Department of Physics, Presidency University,\\
86/1 College Street, Calcutta-700073, INDIA}\\
\emph{$^3$Centre for Theoretical Studies, Indian Institute of Technology,\\
Kharagpur-721302, INDIA}\\ 
\vspace{.4cm}                                             
E-mail:$ ^* $ ranajay95@gmail.com \\
E-mail:$ ^+ $ muktish.physics@presiuniv.ac.in\\
E-mail:$ ^\dagger $abyaya93@gmail.com
         
\end{center}

\vspace{1.5cm}

\textbf{Abstract:} We report the results of the magnetisation reversal in Ising ferromagnet having thermal and field gradients by Monte Carlo simulation. We have studied the distribution of reversal times for different values of thermal and field gradients and compared the results with those obtained for uniform temperature. The movement of the domain wall of distinct domains and the growth of roughness of domain wall have also been studied statistically. The role of competing thermal and field gradients, in the reversal mechanism, was also studied and a {\it line of marginal competition} was obtained. 

\vspace{2.5cm}


\noindent $^a$Corresponding author.

\newpage

\section{Introduction}

The Ising model is a prototype to study the thermally
activated process like nucleation\cite{rik0,hanna}. In the
paramagnetic region, each spin has almost equal probability of finding itself
in the up or down state. As a result, almost equal number of up and down spins
are available in the system. However, in the ferromagnetic region (below the
critical temperature), the system prefers to remain in either of two states: one state
having positive magnetisation where most of the spins are up, and the other state
having negative magnetisation. For a system with finite size, the 
spontaneous reversal of magnetisation (where the magnetisation changes
sign) occurs due to thermal noise. This phenomenon could be observed if the
system is studied over a prolonged period of time.

This magnetisation reversal has a technological application in designing the
magnetic recording devices. In this recording device, for high recording speed,  
external field driven rapid reversal of
magnetisation is required. Let us briefly mention a few studies, related to the field
driven reversal of magnetisation, in Ising ferromagnets.
The large scale computer simulation in Ising
ferromagnet was studied\cite{mads} to identify the distinct regimes of
single droplets and multi-droplets as a precursor of magnetisation reversal. The problem of 
classical nucleation is still challenging. A significant amount of work has already been
done\cite{rik1,ryu,chen,bosia} in this field. However, all the studied mentioned, so far,
have been done for steady (in time) and uniform (over the space) magnetic field.

The direction of research in the field of nucleation has recently been steered towards
the time dependent magnetic field.
The magnetisation reversal transition was studied\cite{bkc} in 
Ising model driven by
an impulsive magnetic field and a phase boundary has been drawn. Recently, the
nucleation and magnetisation reversal have been studied\cite{ma1} 
in Ising ferromagnet driven by
a field spreading in time. Here the competing effects of reversal time and
the rate of spreading have been discussed. The effects of the rate of changing the
temperature on the reversal time was investigated\cite{rikvold}, recently. The magnetisation
switching in classical Heisenberg model for small ferromagnetic particles
was studied extensively\cite{uli}. The experimental evidences of magnetisation switching has also been
reported\cite{expt1}. In these cases, the magnetic field was uniform over
the space.

What happens when the applied field
has a spatial variation, can be a triggering question.
Very recently, the magnetisation reversal in Ising ferromagnet 
by a field having a spatial gradient,
was studied
\cite{abyaya} by extensive Monte Carlo simulation. In that case, the 
temperature of the system was uniform. In this paper, we have studied the
magnetisation reversal, in Ising ferromagnet, where the system has a
nonuniform spatial distribution of temperature. 

In this article, we report the results of our studies on systems having
non-uniform spatial distribution of temperature alongwith the simultaneous
presence of spatially distributed non-uniform external magnetic field. The paper is organised
as follows: In the next section, section 2, we have described the model,
the results of the simulation are reported in section 3, the paper ends with
a discussion in section 4. 

\section{Model}

The Hamiltonian of a two dimensional Ising ferromagnet with nearest neighbour interaction and in presence of  magnetic field (which can vary spatially) can be represented as,
\begin{equation}
H=-J\sum_{<i,j>}{S_{i}}.{S_{j}}- \sum_{i}{h(i).S_{i}}
\end{equation}
Here $S_{i}(={\pm}1)$ represents the Ising spin at i-th lattice site. $J (>0)$ is the ferromagnetic interaction strength and $h(i)$ is the external magnetic field at the i-th lattice site.

In the first part of our work, we have considered the uniformly applied magnetic field and 
the temperature has a spatial variation
along the $x$ direction only), instead of the considering the spatially uniform (temperature) cases in the previous studies\cite{mads,ma1,abyaya}. As a result there is a gradient in temperature between the temperature of the left edge and the right edge of the lattice. We have considered the linear variation of the temperature as:
\begin{equation}
T(x)=g*x+c.
\end{equation}
where, $T(x)$ is the temperature of the x-th column of the 
square lattice.
The thermal gradient is defined as, $g={{dT(x)} \over {dx}}$. 
 
 If initially, $T_{l}$ and $T_{r}$ are the assigned temperatures on the left boundary and the right boundary of the lattice respectively, the thermal gradient, $g$ can be calculated as $g=(T_{r}-T_{l})/L$. The constant $c$ (in equation-2) will be equal to $T_{l}$. 
Here all the magnetic fields are measured in the units of $J$.

In this study, we have fixed $T_{l}=1.4 J/k_{B}$ and have varied $T_{r}$ at values less than 1.4 J/$ k_{B}$ to set the desired thermal gradient, in the system.

 In the second part of our study, we have the temperature gradient as described. In addition to that, we have introduced a field gradient in the system. Let $h_{l}$ and $h_{r}$ be the applied magnetic fields on the left and the right boundaries of the lattice respectively. The magnetic field varies linearly between the two boundaries. So 
 the field takes the form:
 \begin{equation}
          h(x)=f*x + d
 \end{equation}
 where, $f={{dh(x)} \over {dx}}$ is the field gradient. We have thus introduced both gradients of the field and temperature in this part of our study. Needless to say that $d=h_l$ (in equation-3).
 
 We have kept $h_{r}$ fixed at -0.5 and varied $h_{l}$ to get the required field gradients.        
 
\section{Calculation}

In this simulation, we start with a two dimensional lattice of size $L{\times}L$. We use {\bf open boundary conditions} in all directions. Initially, we have coinsidered, $S_{i} =+1 {\forall} ~i$ (perfectly ordered state). Next, at each time step, we follow random updating method. We select any site randomly. Calculate the change in energy $H$ (equation-1) for that spin to flip ($S_{i}{\rightarrow}-S_{i}$). Let that energy be ${\Delta H}$. The  probability that the randomly selected spin will flip, is given by the Metropolis formula \cite{binder}:
$P={\rm Min}(1,\exp{(-{\Delta}H/k_{B}T(i,j))})$. 
The temperature $T(i,j)$ (at any site $(i,j)$) is measured in the unit $ J/k_{B} $, 
where $k_{B}$ is the Boltzmann constant.

The scheme of the simulation can be described briefly as follows: A  
 random number (uniformly distributed between 0 and 1), $r$ , is called. If this random number $r$ is less
than or equal to Metropolis flipping probability $P$  then flipped the selected spin. In this way,
$L^2$ such spin
were flipped in random updating scheme. This $L^2$ number of spin flips is taken to be a single
time step and defined as one time unit (Monte Carlo Step per Spin or MCSS).

For our work, we have chosen $L=300$. The reason behind this choice is a compromise
between the affordable computational time and to have the clear observation of distribution of
metastable lifetimes. 

\vskip 1cm

\section{Results}
 \subsection{Presence of Thermal Gradient Only }

\begin{figure}[h]
\begin{center}
\begin{tabular}{c}
a
\resizebox{8cm}{5cm}{\includegraphics[angle=0]{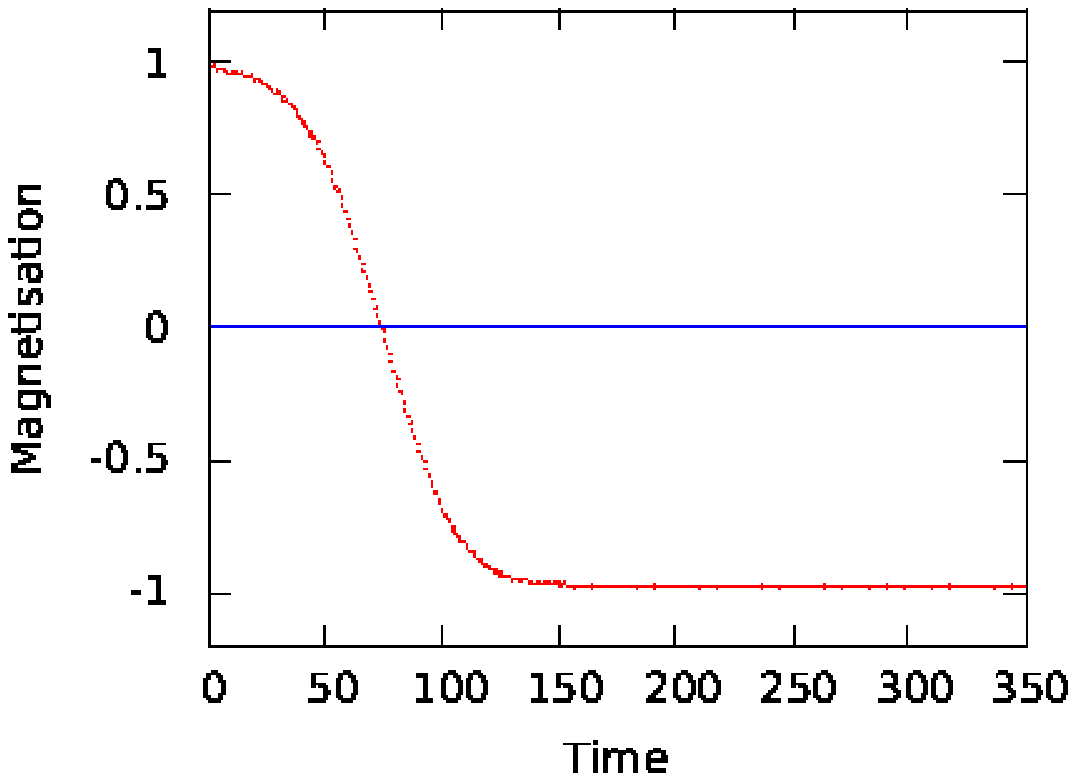}}
b
\resizebox{8cm}{5cm}{\includegraphics[angle=0]{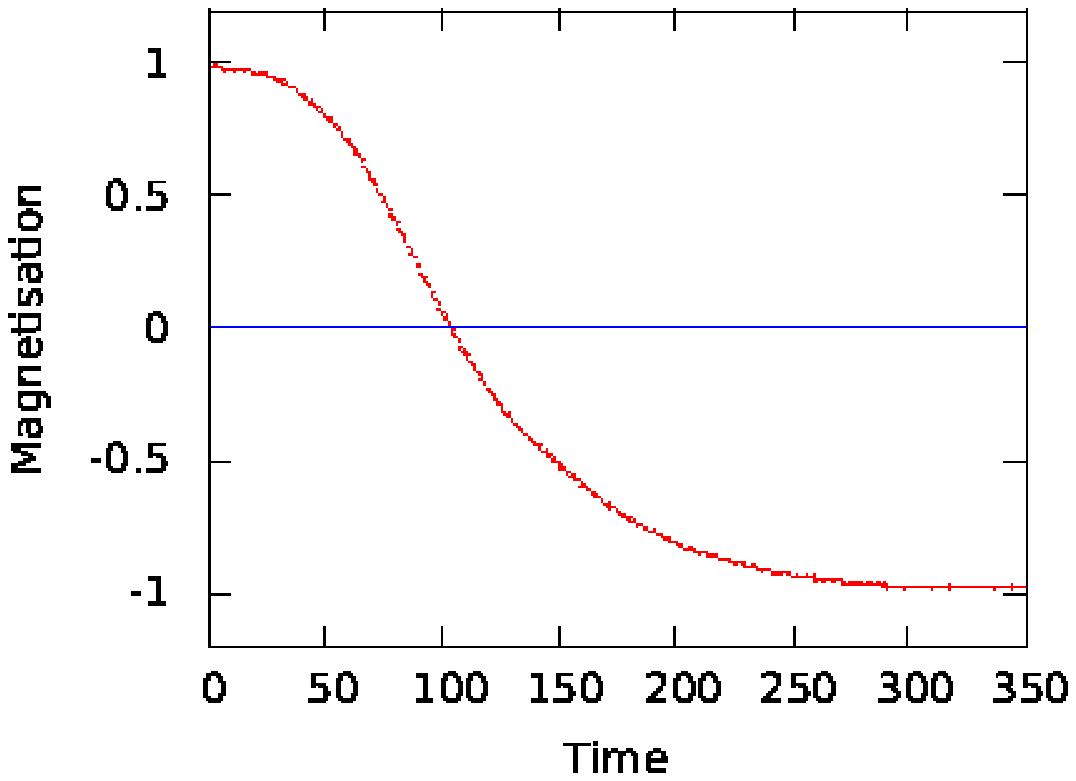}}

\\
c
\resizebox{8cm}{5cm}{\includegraphics[angle=0]{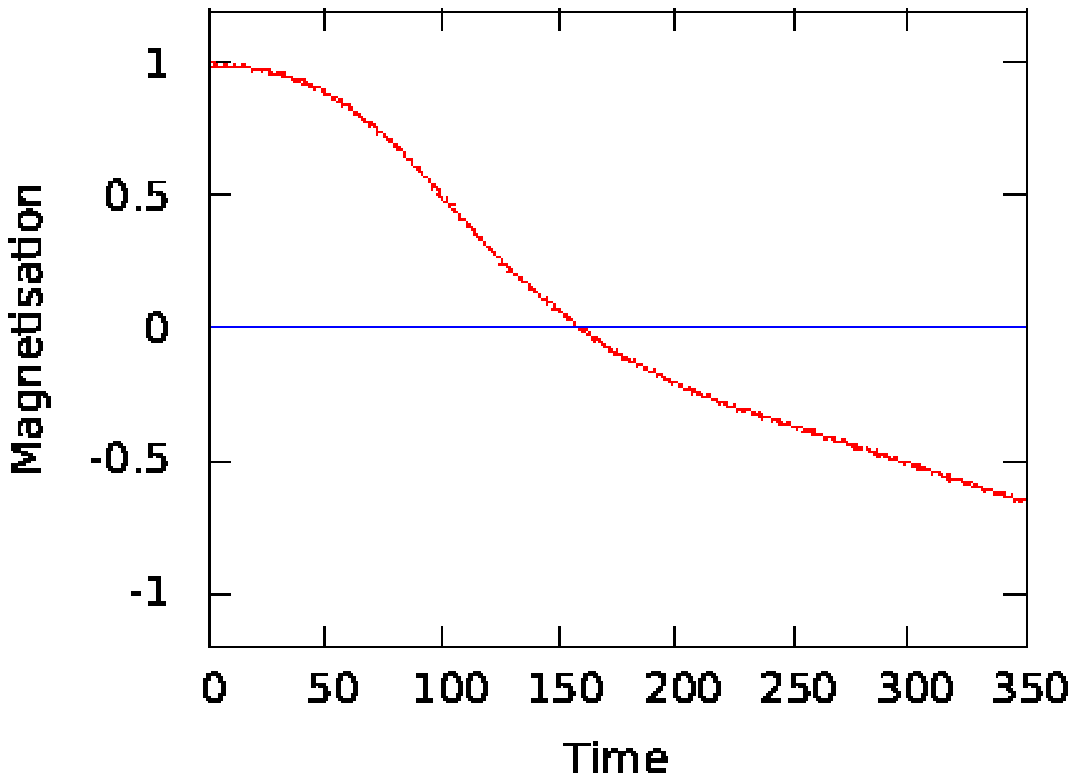}}
d
\resizebox{8cm}{5cm}{\includegraphics[angle=0]{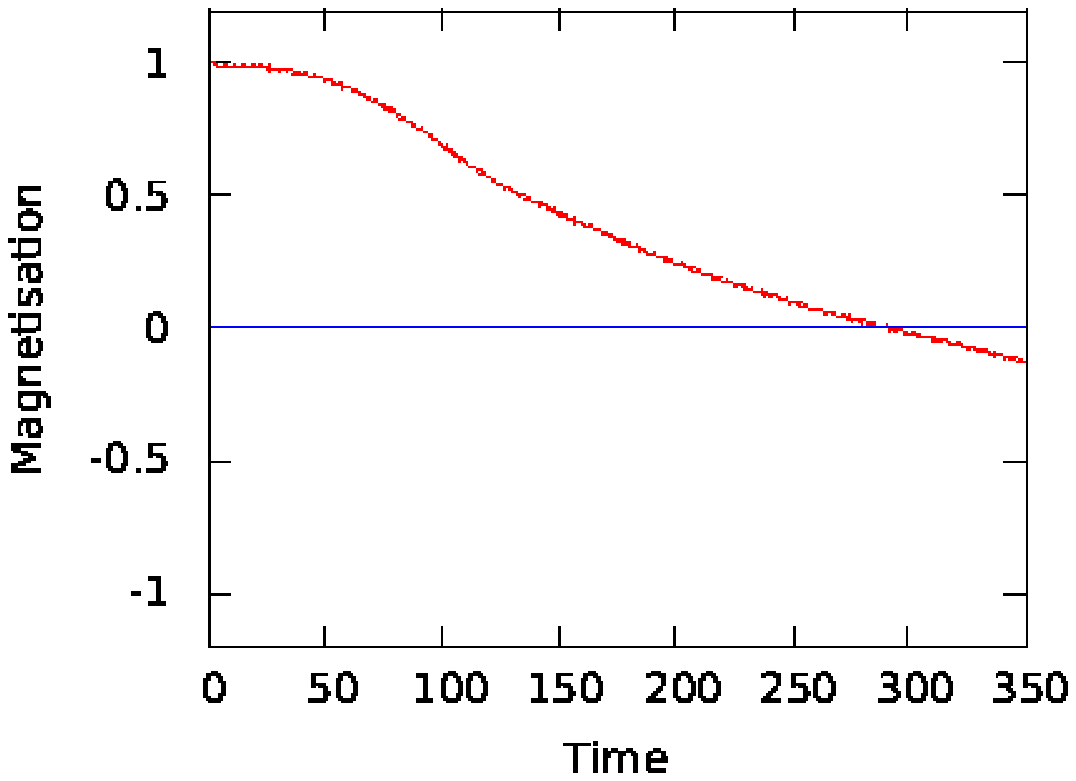}}

          \end{tabular}
\caption{
Variations of magnetisation with time. {\bf Starting from top}  (a) The system at uniform temperature $T(i,j)=1.4$ (b) The system with $T_{l}$=1.4 and $T_{r}$=1.2  (c) The system with $T_{l}$=1.4 and $T_{r}$=1.0 (d) The system with $T_{l}$=1.4 and $T_{r}$=0.8, where $T_{l}$ and $T_{r}$ are the temperatures on left boundary and right edge of the lattice respectively. Temperatures are measured in the units of $J/k_{B}$.
}
\label{fig:reversal}
\end{center}
\end{figure}

 Initially,the system is kept in a fully magnetised state
($S_i=+1{\forall} ~i$). There is an external magnetic field applied in the opposite (to the direction of initial magnetisation) direction. Then the minimum time taken by the system to reach a negative magnetisation state from the completely ordered initial state of positive magnetisation, in the presence of the applied external magnetic field is called the reversal time or the metastable lifetime of the ferromagnetic system. Fig-\ref{fig:reversal}a shows the variation, of the magnetisation of such a system, with time.

We have studied the reversal of magnetisation of the system having the thermal gradient
(but in the presence of uniform magnetic field). 
For different values of the thermal gradient, such variations of 
magnetisation with time, are shown in Fig-\ref{fig:reversal}a,b,c and d.

\begin{figure}[h]
\begin{center}
\begin{tabular}{c}
a
\resizebox{8cm}{6cm}{\includegraphics[angle=0]{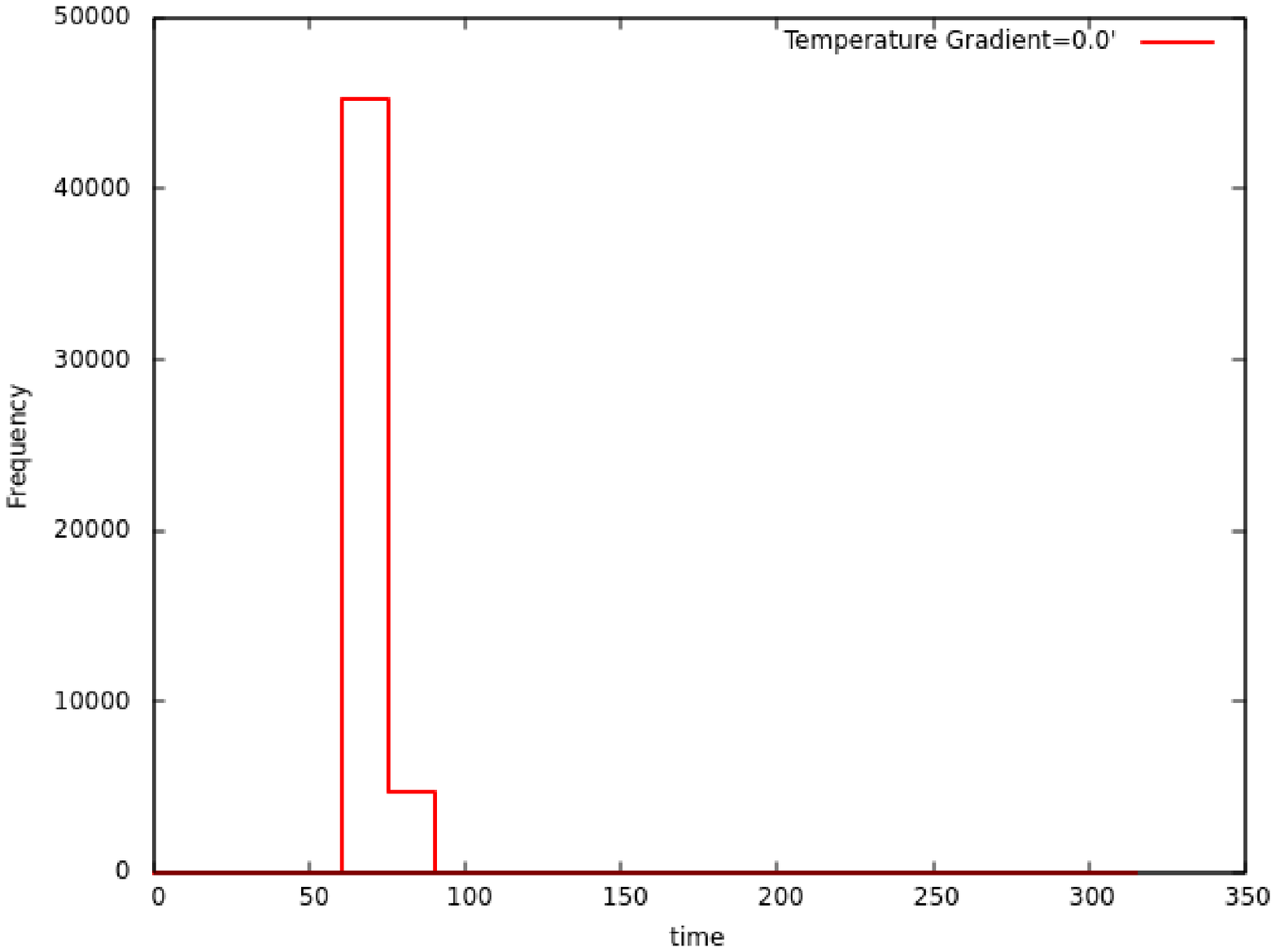}}
b
\resizebox{8cm}{6cm}{\includegraphics[angle=0]{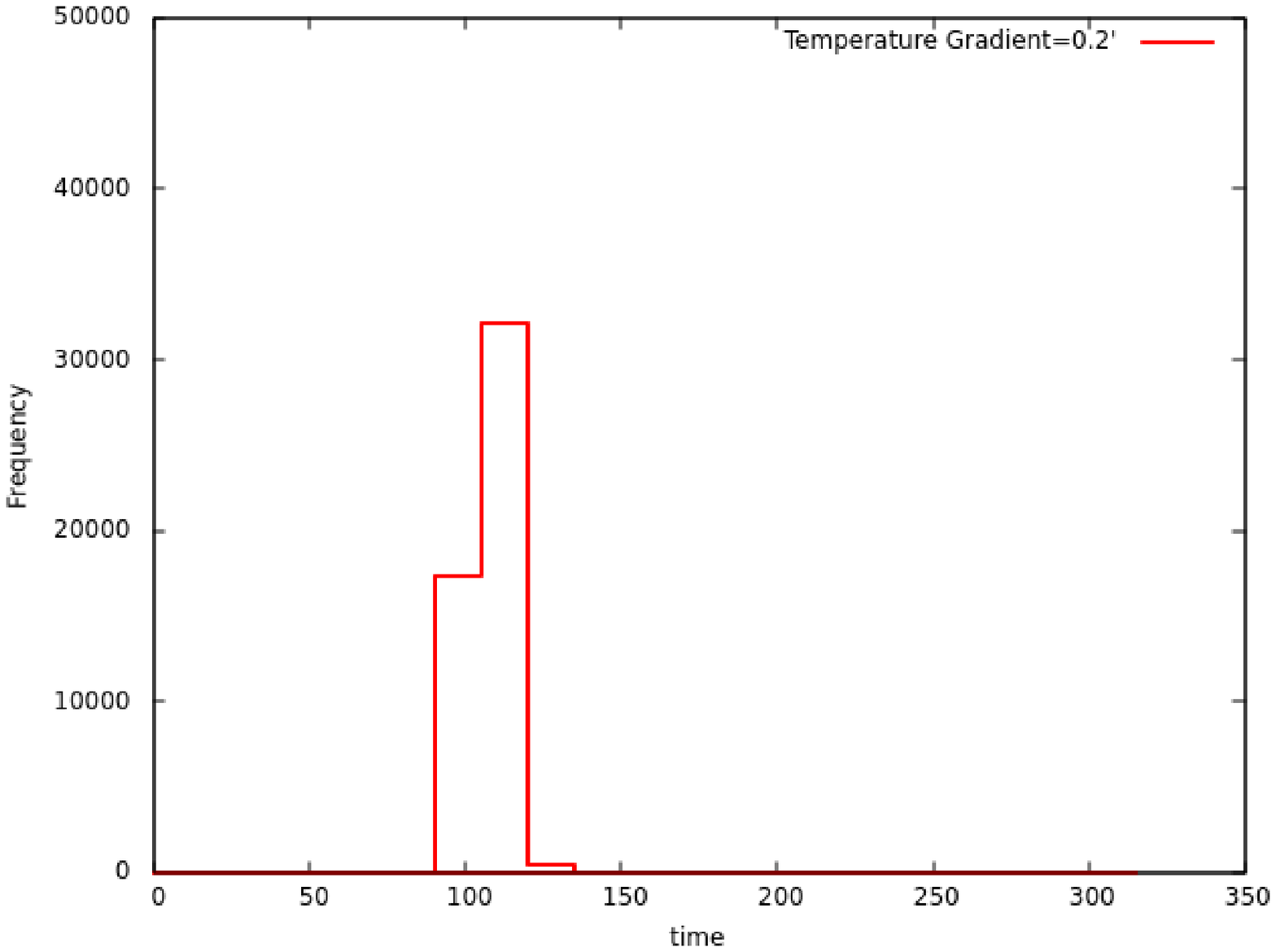}}
\\
c
\resizebox{8cm}{6cm}{\includegraphics[angle=0]{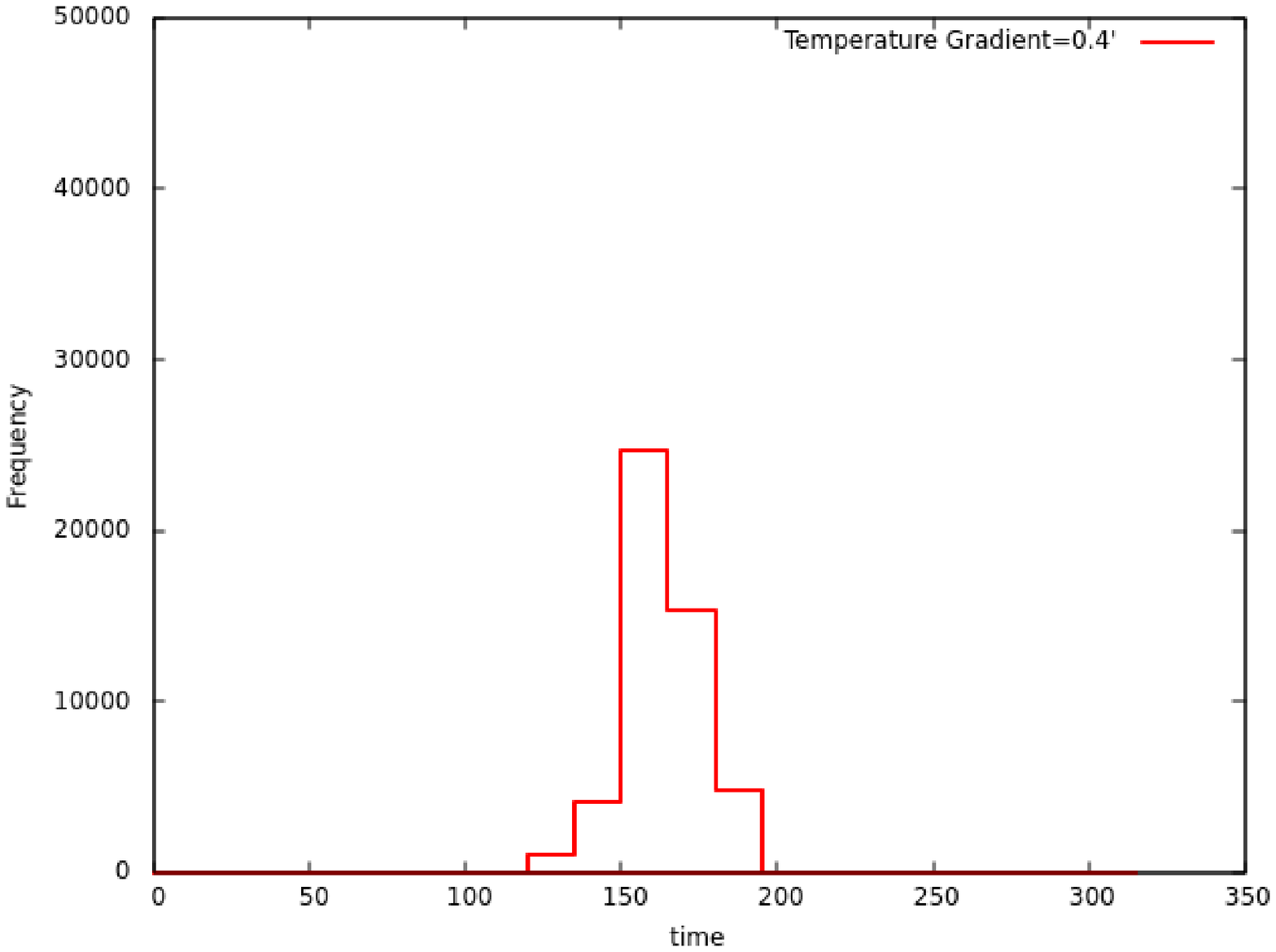}}
d
\resizebox{8cm}{6cm}{\includegraphics[angle=0]{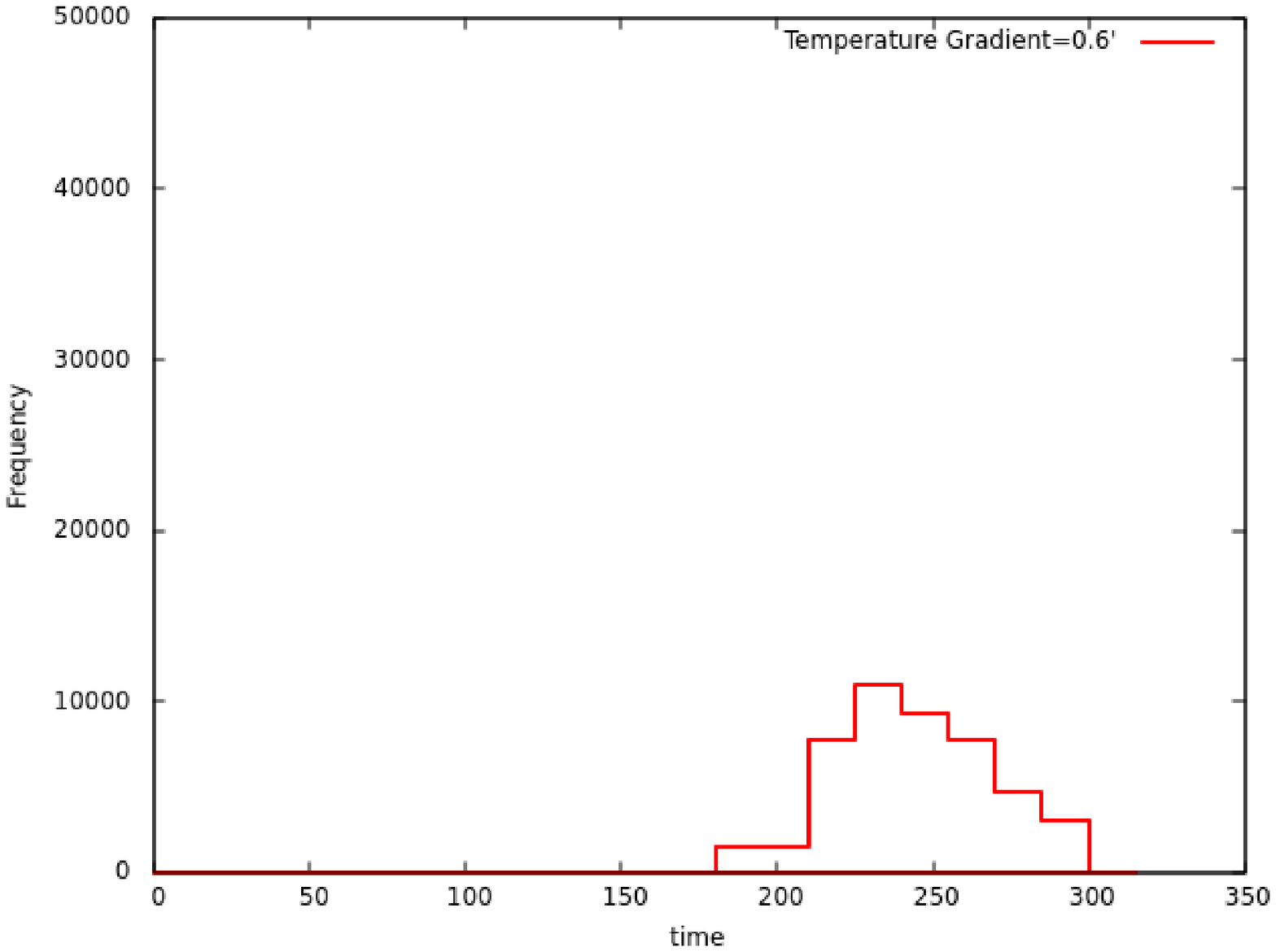}}
          \end{tabular}
\caption{Distributions of reversal time of 50,000 ferromagnetic samples for different thermal gradients. {\bf Top left} (a) The system at uniform temperature $T=1.4$.  {\bf Top right} (b) The system with $T_{l}$=1.4 and $T_{r}$=1.2. {\bf Bottom left} (c) The system with $T_{l}$=1.4 and $T_{r}$=0.0. {\bf Bottom right} (d) The system with $T_{l}$=1.4 and $T_{r}$=0.8. Where $T_{l}$ and $T_{r}$ are the temperatures on left  and right edges of the lattice respectively. Temperatures are measured in the units of $J/k_{B}$.} 
\label{fig:distt}
\end{center}
\end{figure}

\begin{figure}[!tbp]

  \centering
    a
    \includegraphics[scale=0.5]{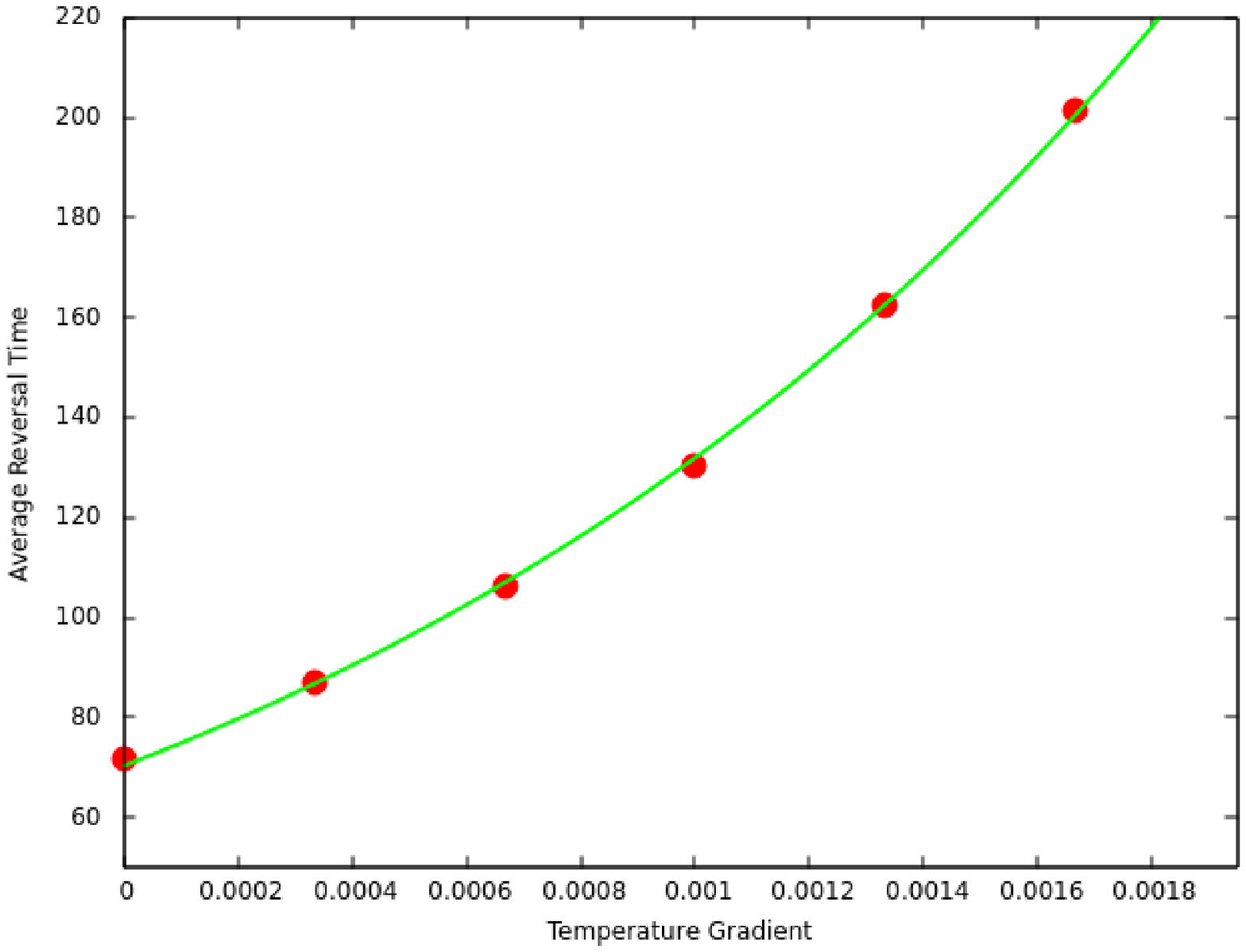}
\\
    b
    \includegraphics[scale=0.5]{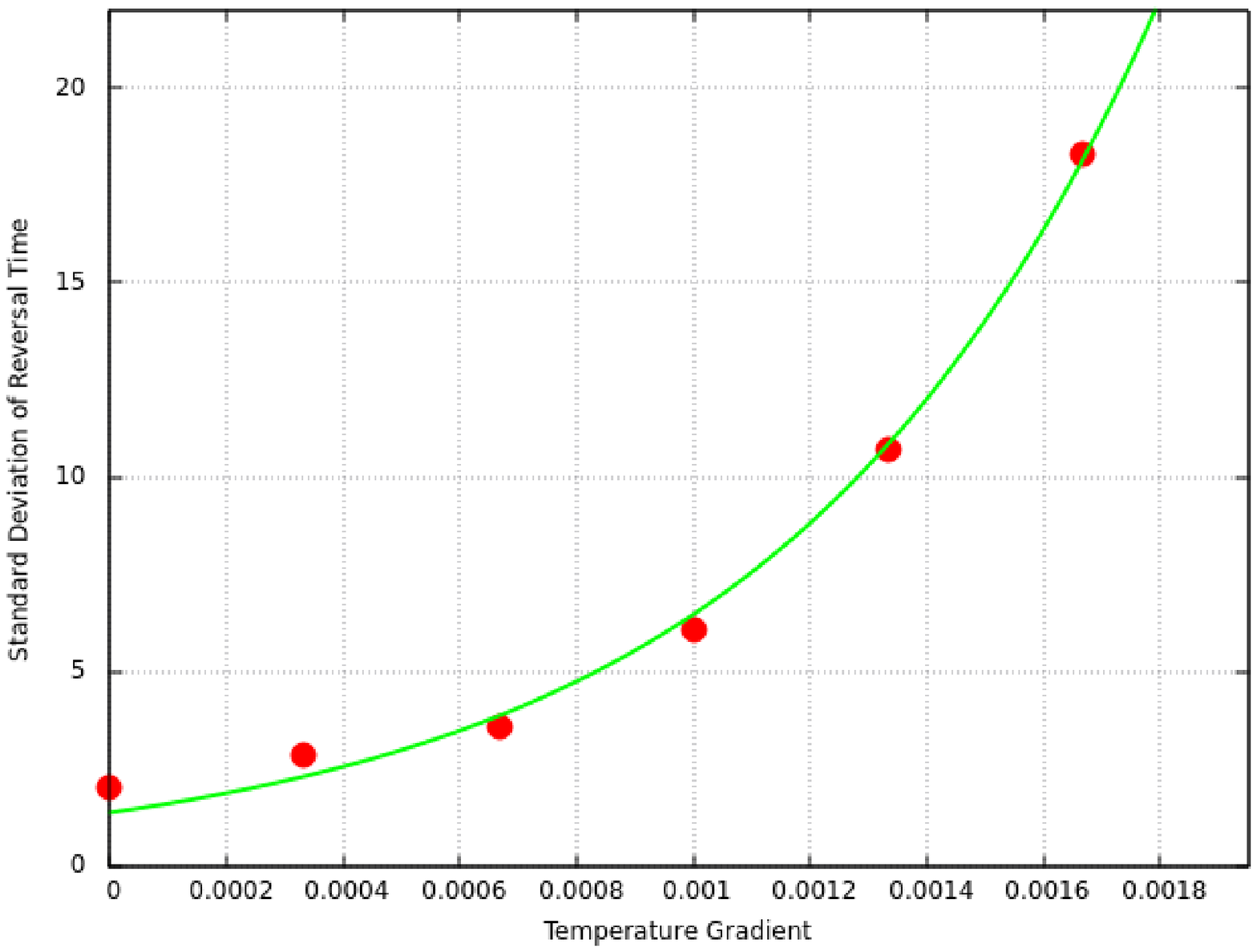}

\caption{Variations of average reversal time and standard deviation of reversal time with thermal gradient. {\bf Starting from top}  (a) Average reversal time varies with gradient as $y \sim {\rm exp}(cx)$ (b) Standard deviation of reversal time varies with gradient as $y \sim{\rm exp}(c'x)$. Thermal gradients are measured in the units of $J/k_{B}$.}
\label{fig:tfun}
\end{figure}

It is observed that as the thermal gradient of the system increases, the reversal time of the system also increases. The distributions of reversal times, obtained over 50,000 such different samples, are shown in Fig-\ref{fig:distt} for different values of thermal gradient (where the external magnetic field remaining the same $h=-0.5$). The distribution is unimodal and the most probable reversal time increases as the thermal
gradient increases. Moreover, the distributions get wider for higher values of the thermal gradient.
We have found that the average reversal time (Fig-\ref{fig:tfun}a) and 
its standard deviation   
(Fig-\ref{fig:tfun}b) increases exponentially as the thermal gradient increases when the 
magnetic field is uniform.  

\begin{figure}[h]
\begin{center}
\begin{tabular}{c}
a
\resizebox{7cm}{6cm}{\includegraphics[angle=0]{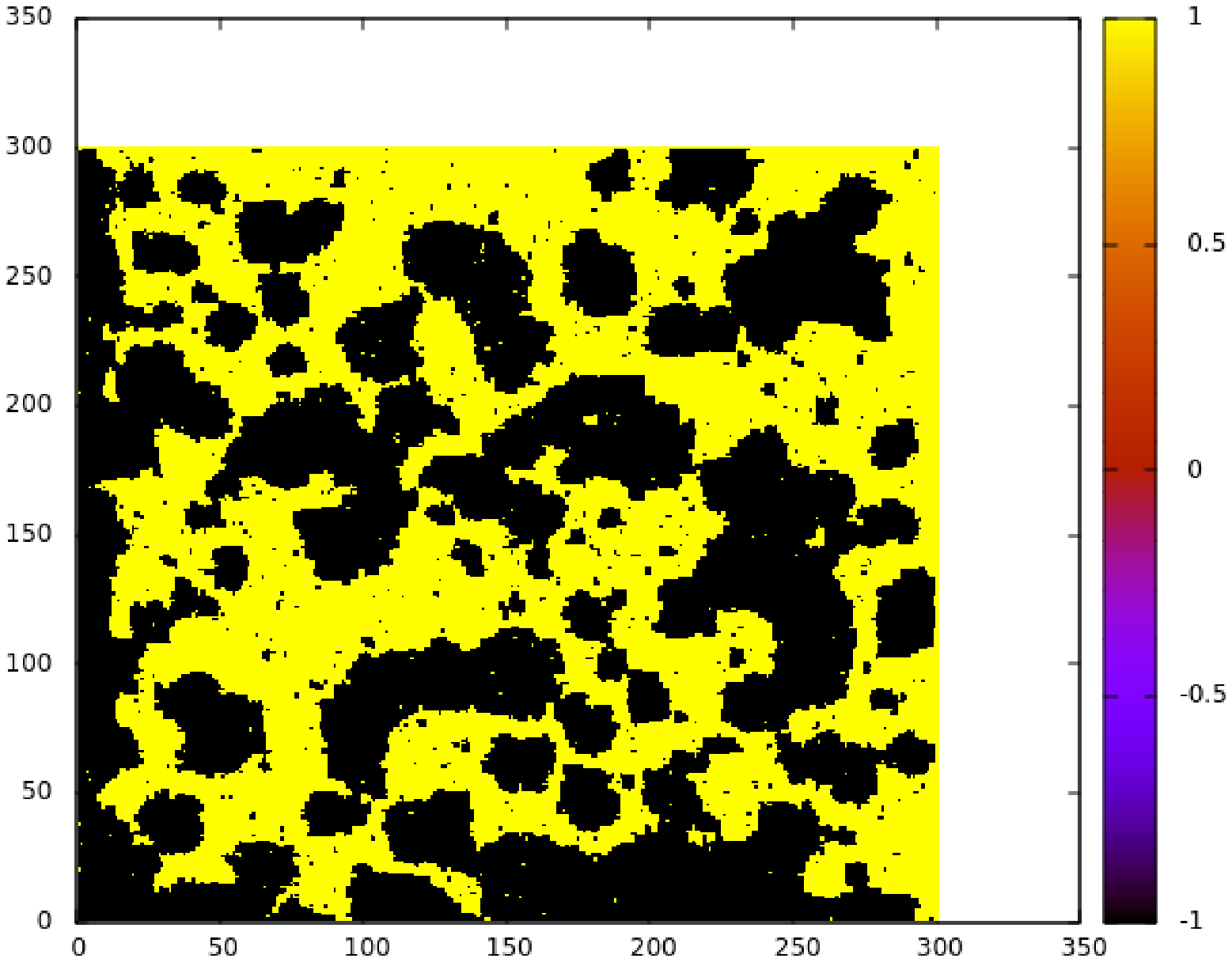}}
b
\resizebox{7cm}{6cm}{\includegraphics[angle=0]{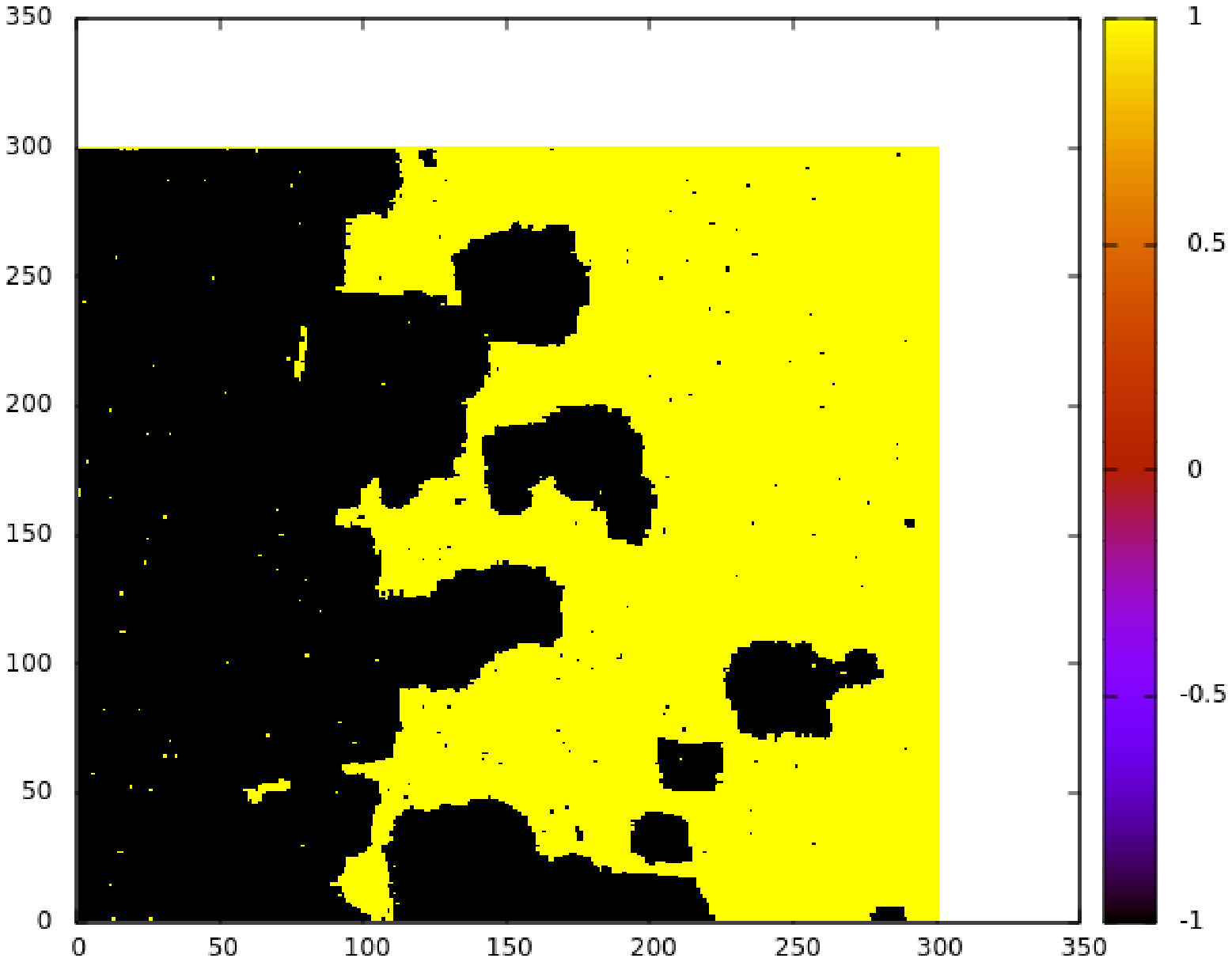}}
\\
c
\resizebox{7cm}{6cm}{\includegraphics[angle=0]{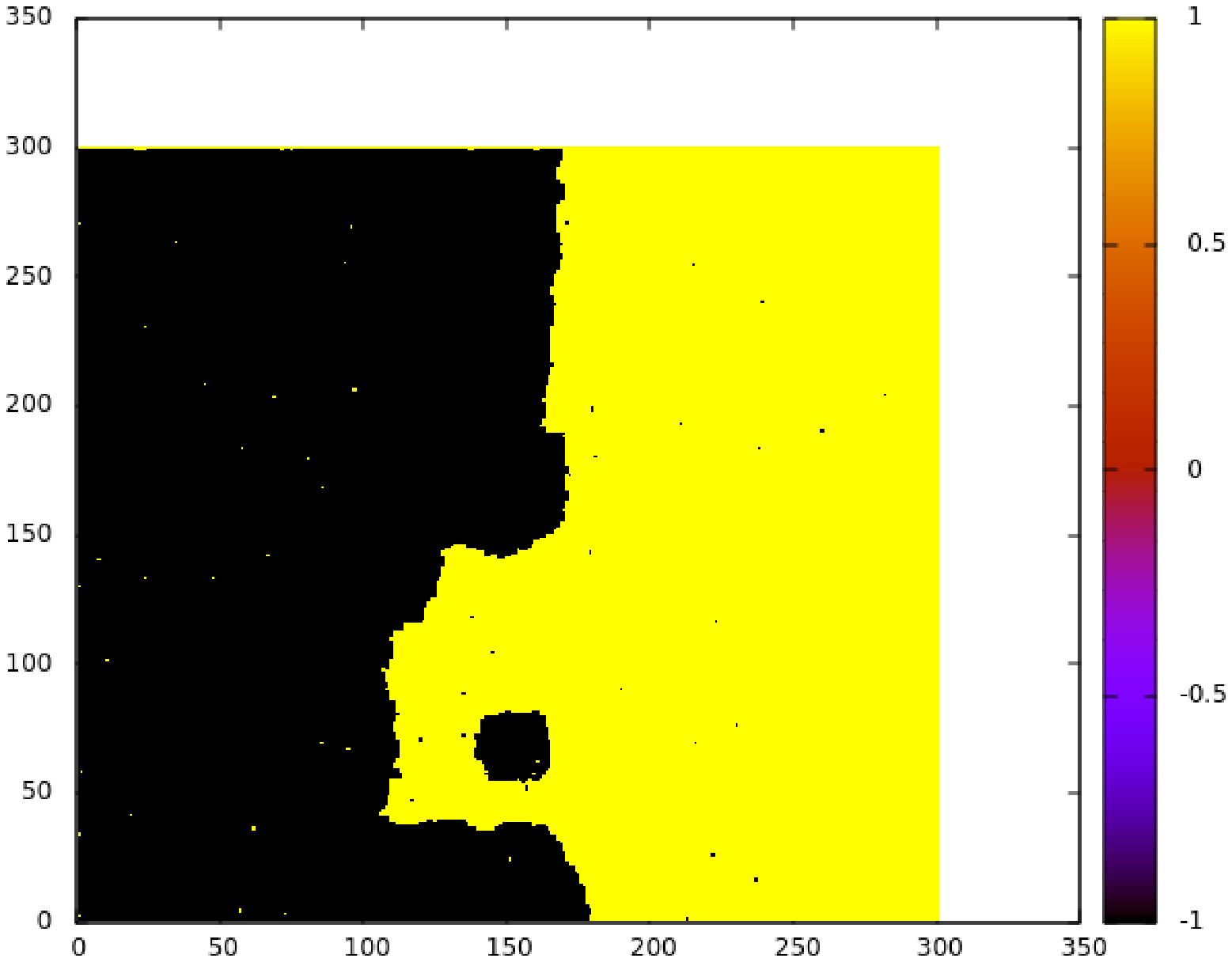}}
d
\resizebox{7cm}{6cm}{\includegraphics[angle=0]{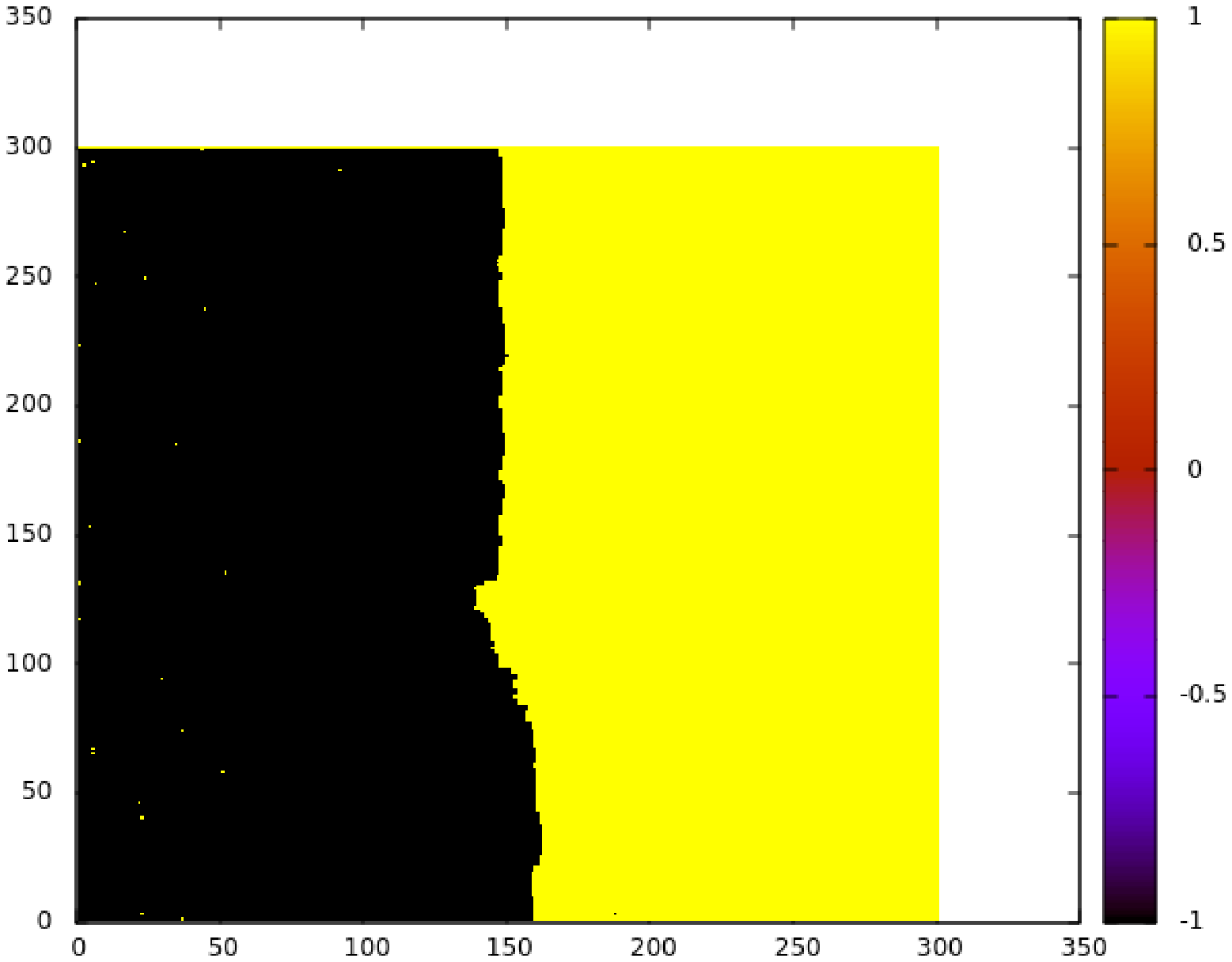}}
          \end{tabular}
\caption{
	Snapshots of the spin configurations at reversal time for different sets of temperature gradients {\bf Top left} a) The system at uniform temperature T=1.4.  {\bf Top right} b) The system with $T_{l}$=1.4 and $T_{r}$=1.0. {\bf Bottom left} c) The system with $T_{l}$=1.4 and $T_{r}$=0.7. {\bf Bottom right} d) The system with $T_{l}$=1.4 and $T_{r}$=0.2, where $T_{l}$ and $T_{r}$ are the temperatures on left  and right edges of the lattice respectively. Temperatures are measured in the units of $J/k_{B}$. Here, the black dots 
represent down spins.
}
\label{fig:morph4}
\end{center}
\end{figure}

The morphology (i.e, the snapshot of the spin configurations of the lattice at a particular instant of time) of the samples are studied by observing the microstates (at the time of reversal) for different values of the temperature gradient. As Fig-\ref{fig:morph4}a shows, in the presence of uniform temperature and field, the lattice morphology at the time of reversal, shows the formation of multiple droplets
(small clusters of down spins marked by black dots), distributed almost homogeneously throughout the lattice. However, in the presence of a thermal gradient, keeping the field constant, instead of the homogeneous distribution of the clusters, droplets of down spins tend to assemble near the boundary having higher temperature. And upon increasing the gradient, we observed that, at the time of magnetisation reversal, a rough interface (domain wall) forms, which separates the regions or domains of positive and negative spins(Fig-\ref{fig:morph4}b and Fig-\ref{fig:morph4}c). The roughness of the interface was observed to decrease (or the interface gets smoother)
as the thermal gradient increases. As Fig-\ref{fig:morph4}d shows, for a high value of the thermal gradient, the lattice morphology shows two prominent domains of up and down spins, with a smooth interface.  

To sketch the shape of the interface (a curved line in two dimensions), we employed the following
method: On a particular row, for each spin on that row, we checked the values of its 10 nearest neighbours on both sides. From Fig-\ref{fig:morph4}d, we can say that, at any position on interface, almost all the spins to its left will be down spins (as the temperature of the left edge of the lattice is higher) and almost all the spins to its right are up spins. When for a particular spin (in a particular row), both the number of up spins among its 10 neighbours (to its left) and down spins among its 10 neighbours (to its right) are greater than or equal to 8, we fix that spin position as the x-coordinate of interface on that row. In this way the positions (the x-coordinate only) of all the 300 rows of the lattice are noted. In this way we have
sketched the shape of the interface.  
The average of all the x-coordinates (or positions) is the measure of averge position of the interface
(or domain wall). And the variance of all these x-coordinates may act as a measure
of roughness of the domain wall.

\begin{figure}[h]
\begin{center}
\begin{tabular}{c}
a
\resizebox{7cm}{6cm}{\includegraphics[angle=0]{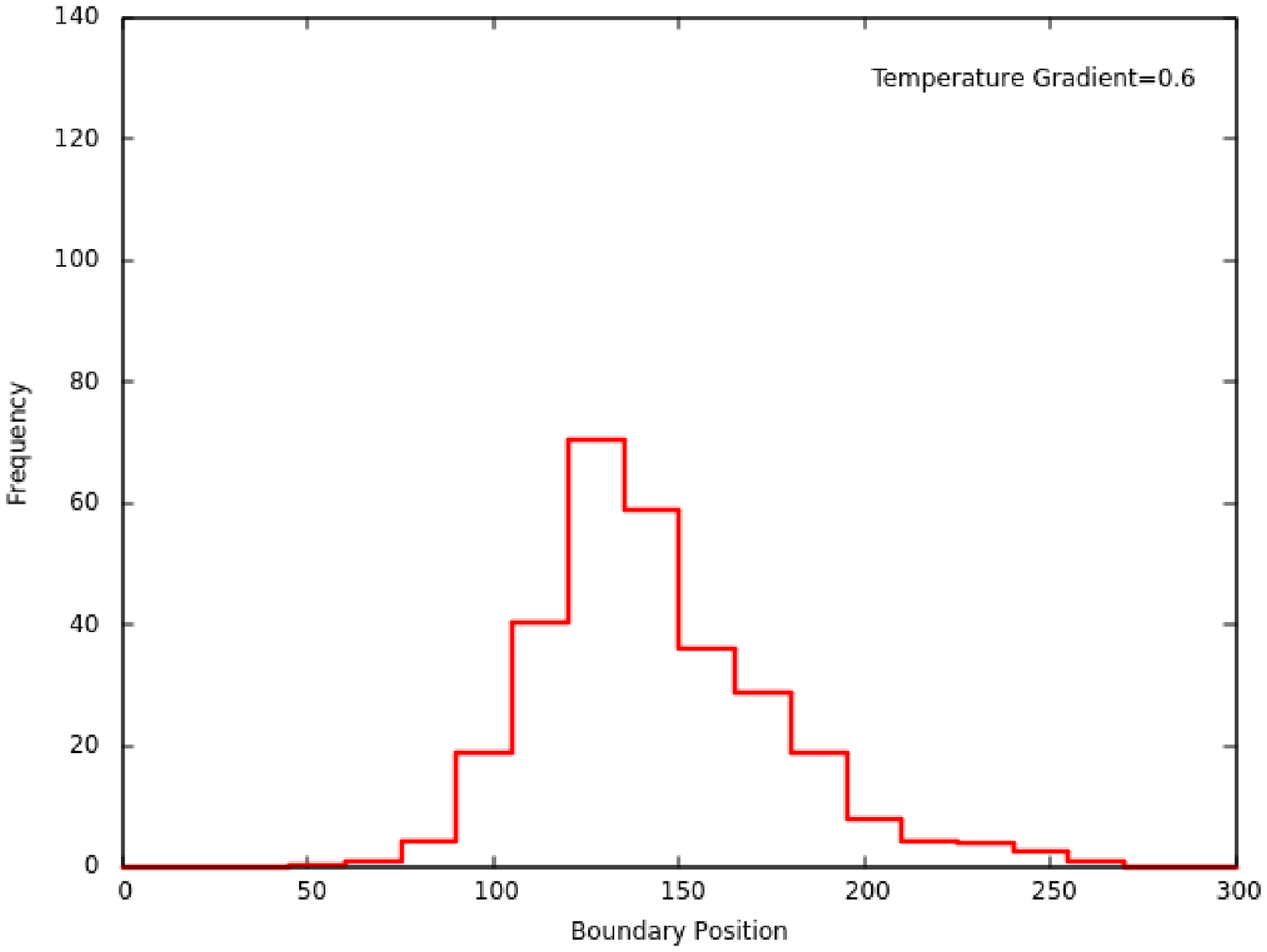}}
b
\resizebox{7cm}{6cm}{\includegraphics[angle=0]{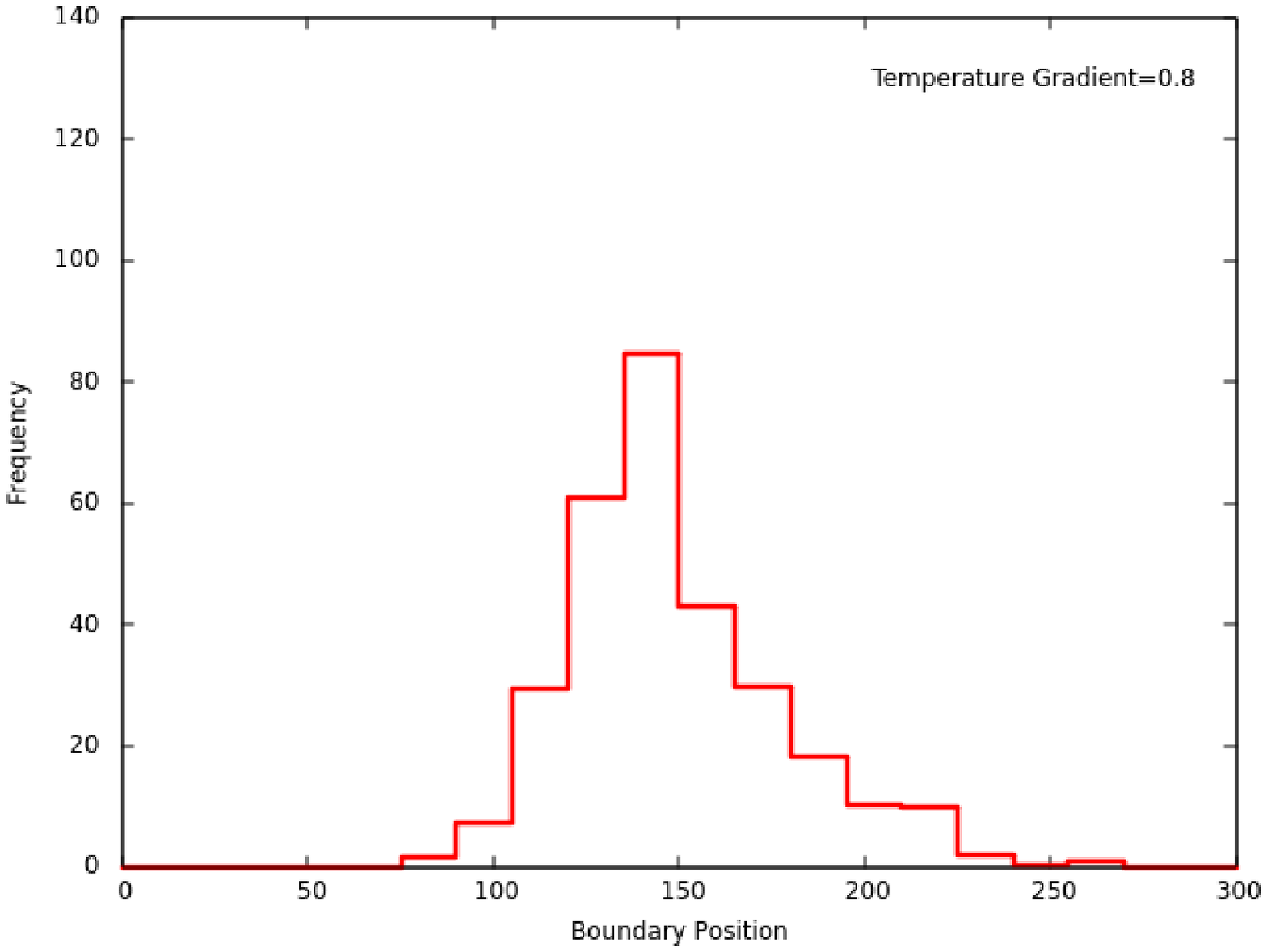}}
\\
c
\resizebox{7cm}{6cm}{\includegraphics[angle=0]{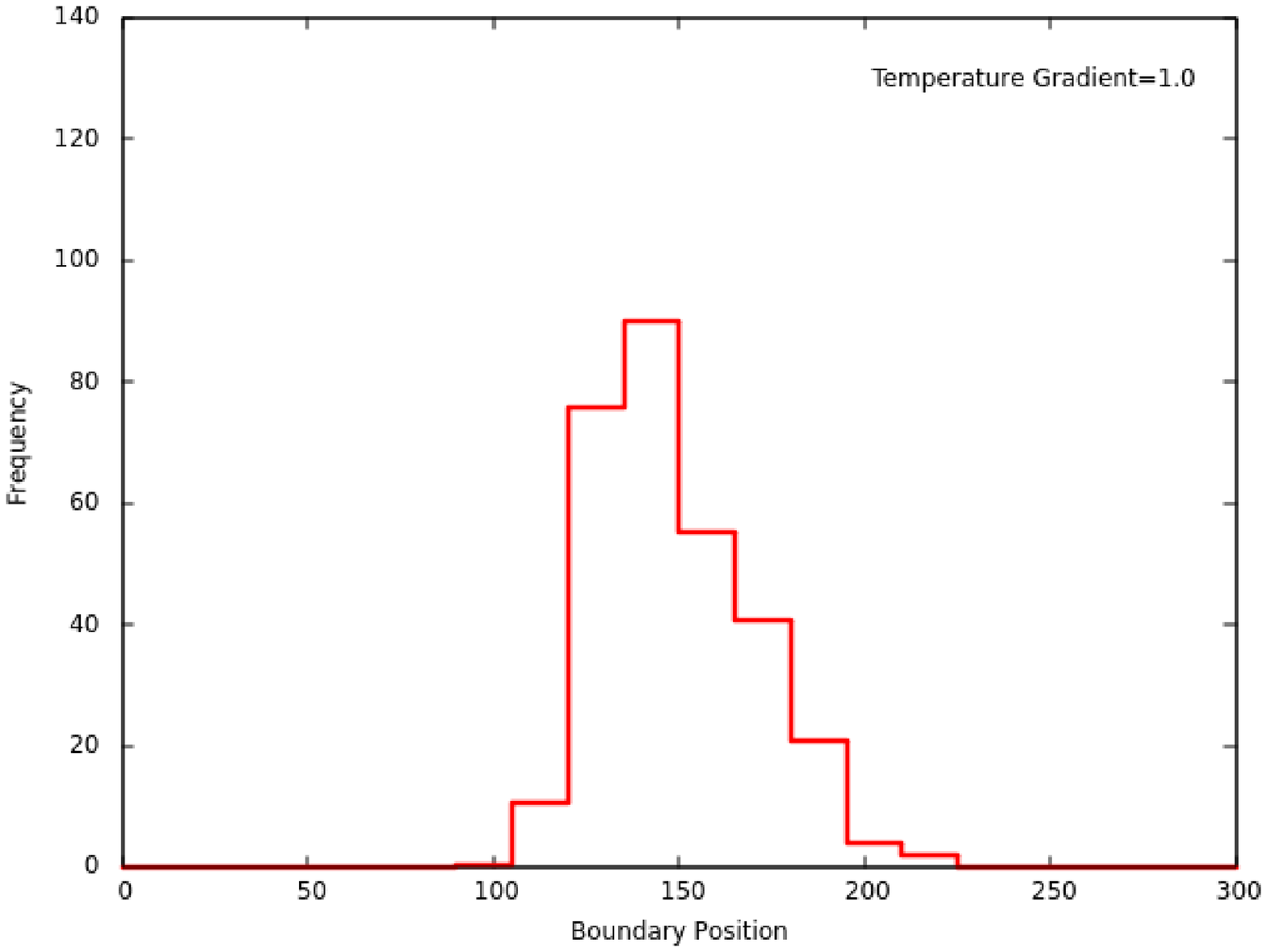}}
d
\resizebox{7cm}{6cm}{\includegraphics[angle=0]{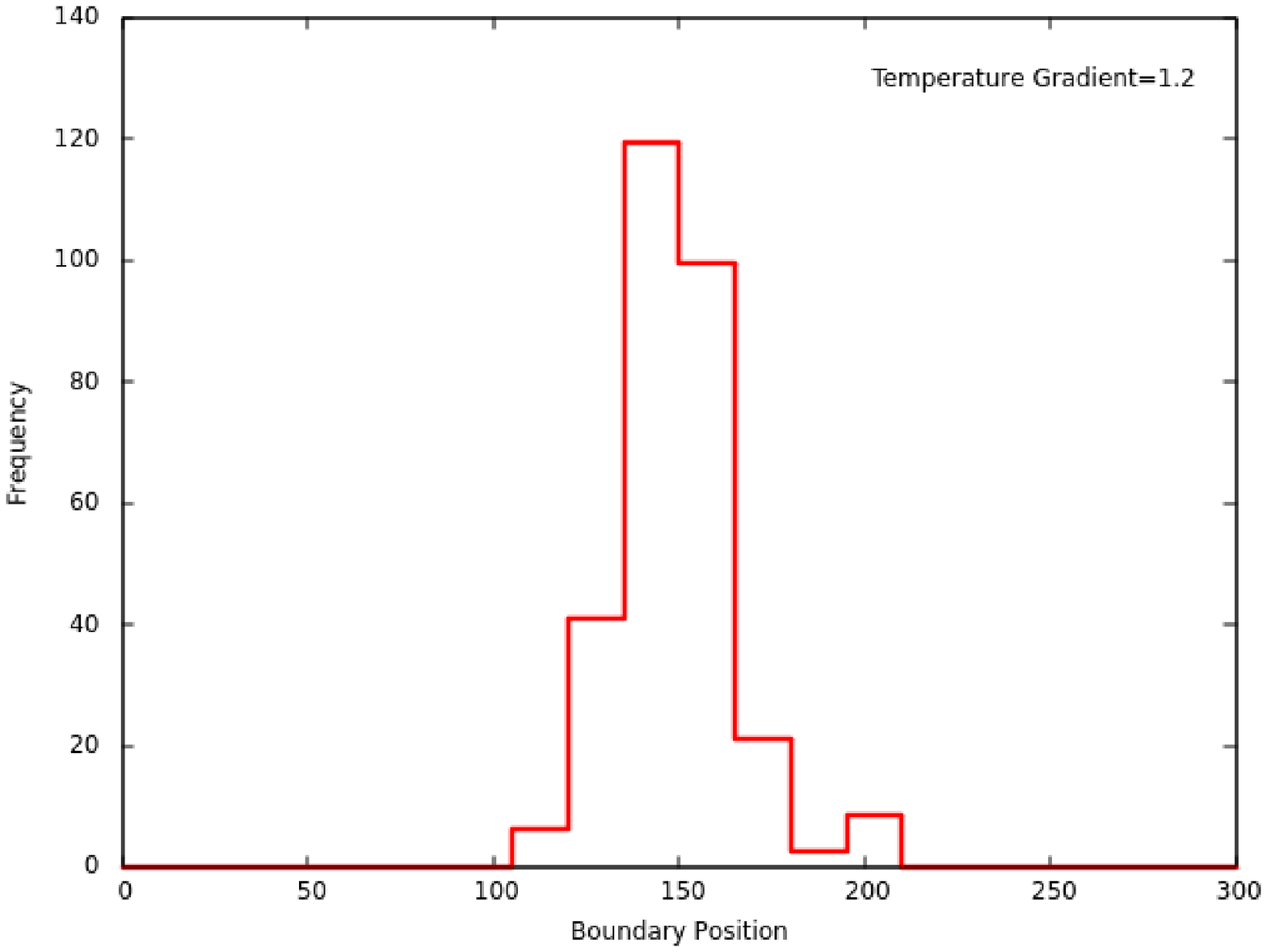}}
          \end{tabular}
\caption{
Distributions of average location of interface (done over 500 samples) for different thermal gradients. {\bf Top left} (a) The system with $T_{l}$=1.4 and $T_{r}$=0.8.  {\bf Top right} (b) The system with $T_{l}$=1.4 and $T_{r}$=0.6. {\bf Bottom left} (c) The system with $T_{l}$=1.4 and $T_{r}$=0.4. {\bf Bottom right} (d) The system with $T_{l}$=1.4 and $T_{r}$=0.2. Where $T_{l}$ and $T_{r}$ are the temperatures on left  and right edges of the lattice respectively. Temperatures are measured in the units of $J/k_{B}$.} 
\label{fig:dist5}
\end{center}
\end{figure}

The unnormalised distribution of interface positions for different gradients (in the form of histogram) is also consistent with the obtained results. If the gradient is increased the width of the distribution decreases (Fig-\ref{fig:dist5}). This observation also provides the testimony to the phenomenon of increasing smoothness of the interface with increasing gradient. One can compare Fig-\ref{fig:morph4} and 
Fig-\ref{fig:dist5}.

\begin{figure}

  \centering
a
    \includegraphics[scale=0.45]{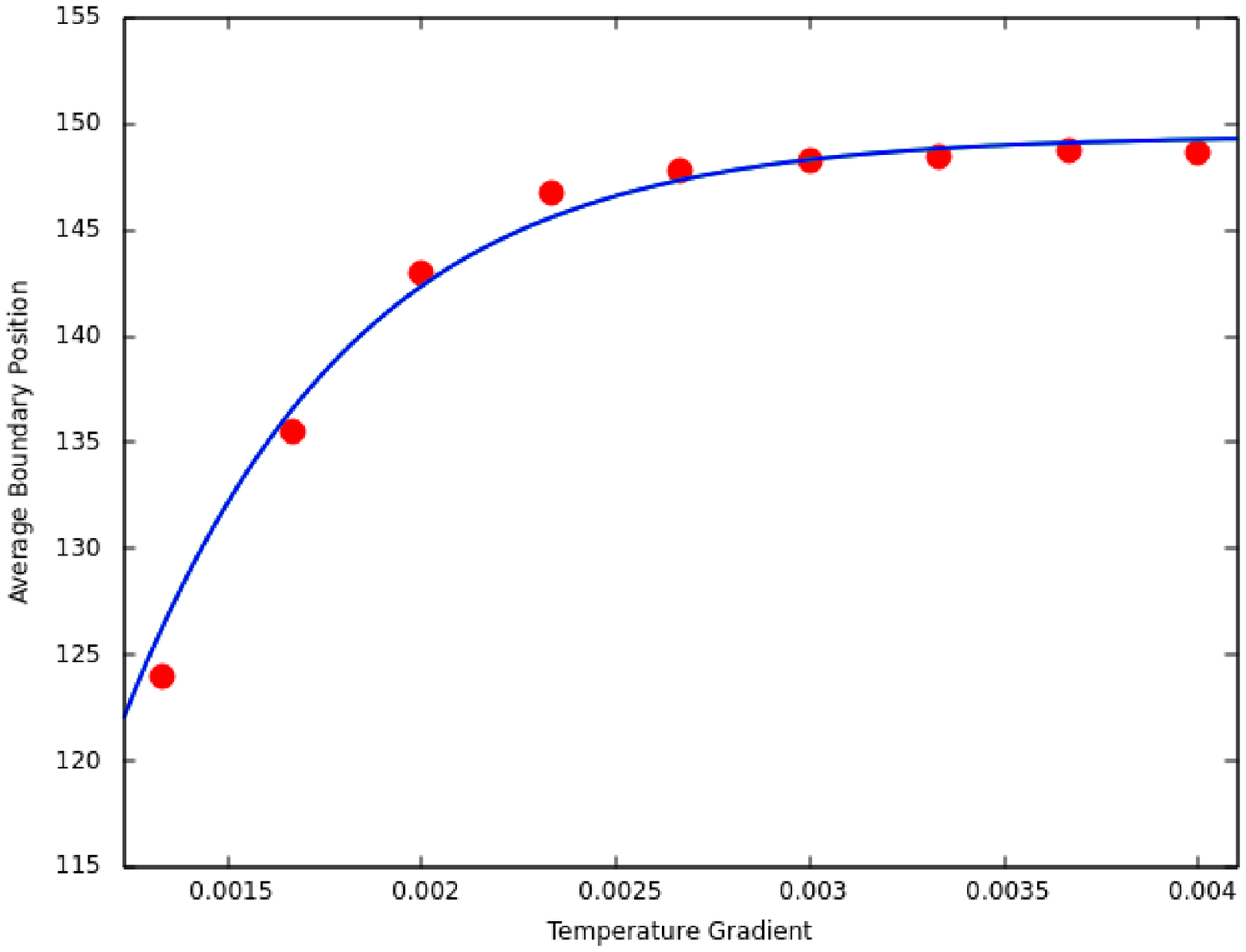}
\\
b
    \includegraphics[scale=0.45]{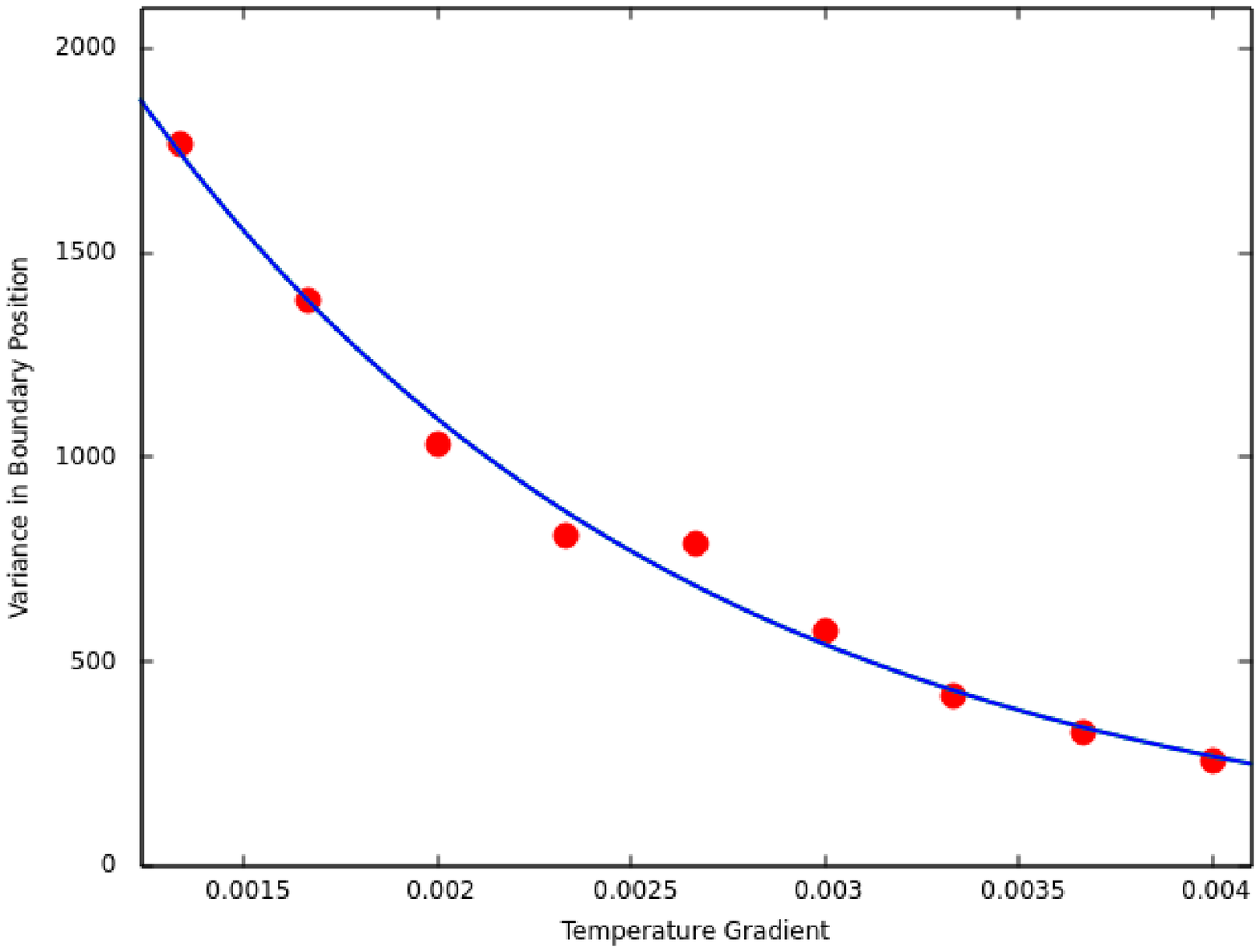}
    
	\caption{Variations of average position of the interface and the variance of interface position with thermal gradient. {\bf Starting from top}  (a) Average interface position varies with gradient as $y=A{\rm tanh}(bx)$ (b) Standard deviation of reversal time varies with thermal gradient 
as $y=C{\rm exp}(-dx)$. Thermal gradients are measured in the units of $J/k_{B}$.}   
   \label{fig:bound} 
\end{figure}

From the plot of the average position of the interface (averaged over 500 different samples) versus 
thermal gradient, it is seen that the average interface position increases upto a certain value and then remains almost fixed to a value near the vertical central line of the lattice (Fig-\ref{fig:bound}a).
This displacement of the mean position of domain wall  
turns out to be a {\it hyperbolic tangent} function of the thermal gradient (Fig-\ref{fig:bound}a).

The plot of the variance in the position 
 of the interface versus the thermal gradient is shown in Fig-\ref{fig:bound}b. It is observed that the variance decreases exponentially (Fig-\ref{fig:bound}b) with field gradient. The variance in boundary position is a good measure of the boundary (of the domain wall) roughness. The observation that the variance decreases with increasing gradient supports the fact that the boundary roughness decreases drastically with increasing gradient. This inference is consistent with our observations of the lattice morphologies of Fig-\ref{fig:morph4} which also showed the decrease in boundary rougness with increasing gradient.

\subsection
{Simultaneous Presence of Thermal and Field Gradients}

Here, we maintain the thermal gradient having the form of equation-2. In addition to that we 
have introduced a field gradient having the form of equation-3.

In our study, $T_{l}$ is kept fixed at 1.4J/$k_{B}$, while $T_{r}$ is varied at values less than 1.4J/$k_{B}$. Also, $h_r$ is kept fixed at -0.5 and $h_l$ is varied at values higher than -0.5. Now, at regions having higher temperature, spins have a higher probability of flipping. Also at regions having higher magnitude of negative field, positive spins have higher probability of flipping. So in assigning a higher magnitude of negative field to the edge having lower temperature and a lower magnitude of negative field to the edge having higher temperature, we have tried to create a situation where the thermal gradient and the field 
gradient compete with each other in their spin flipping ability.

\begin{figure}[h]
\begin{center}
a\begin{tabular}{c}
\resizebox{7cm}{6cm}{\includegraphics[angle=0]{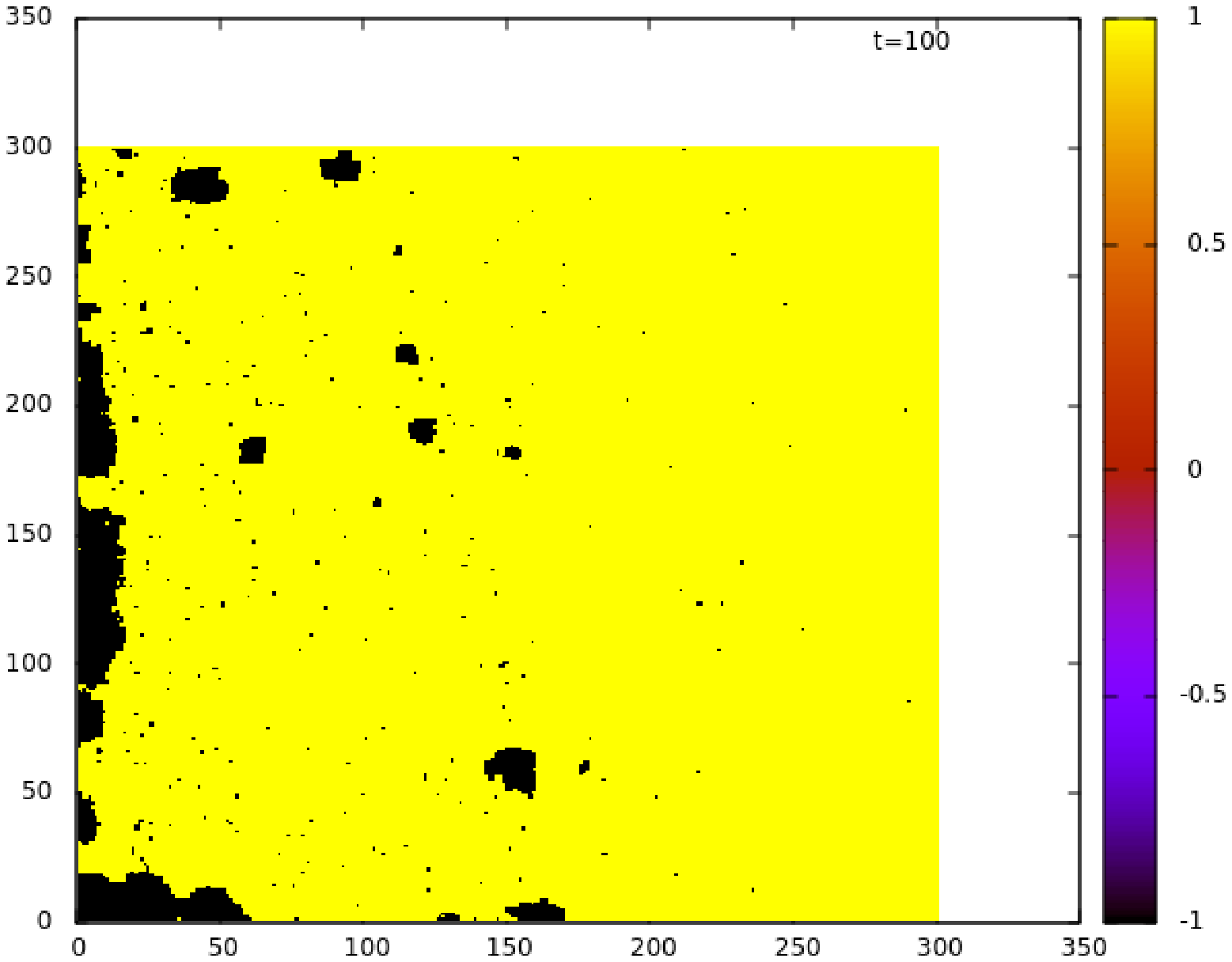}}
b
\resizebox{7cm}{6cm}{\includegraphics[angle=0]{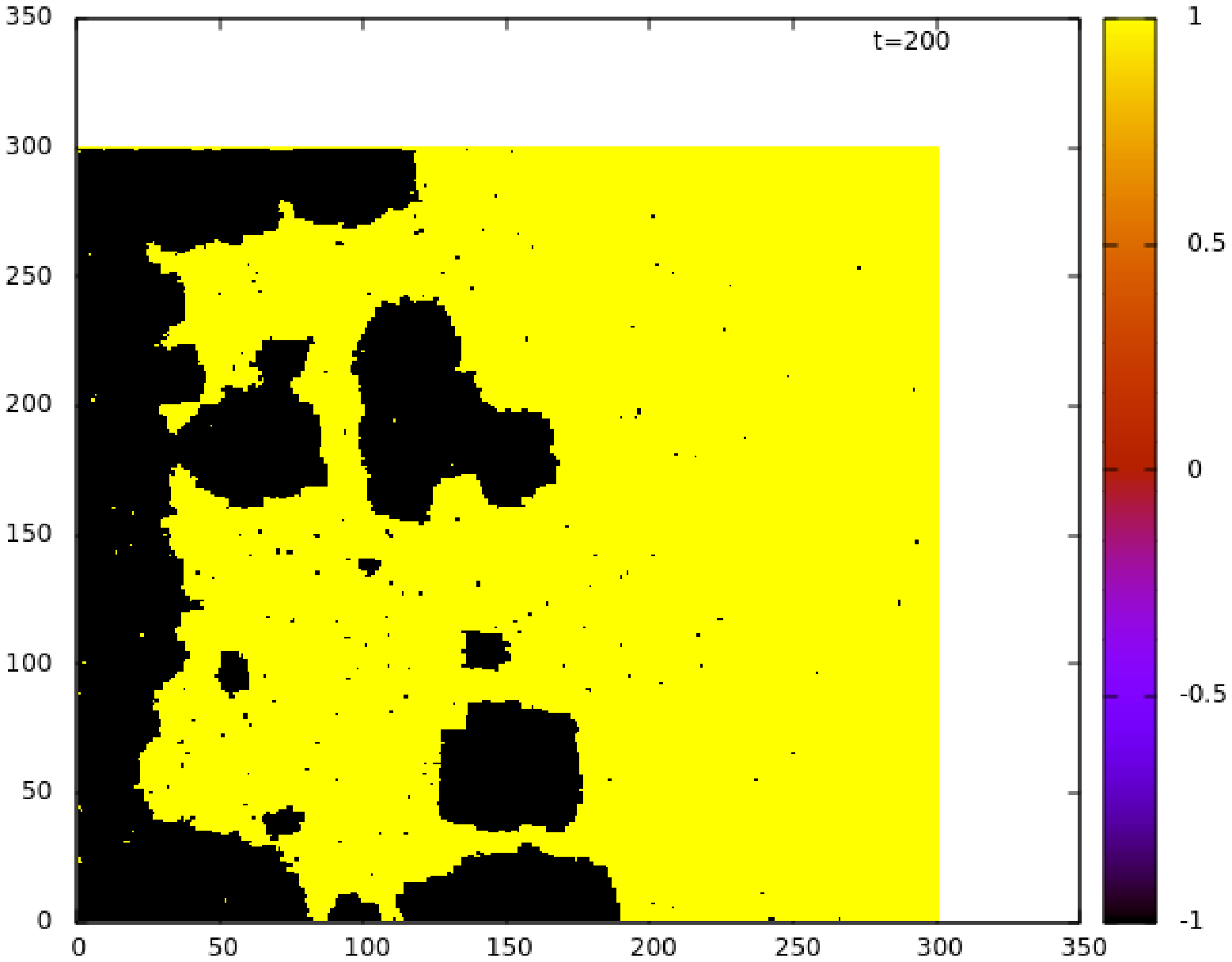}}
\\
c
\resizebox{7cm}{6cm}{\includegraphics[angle=0]{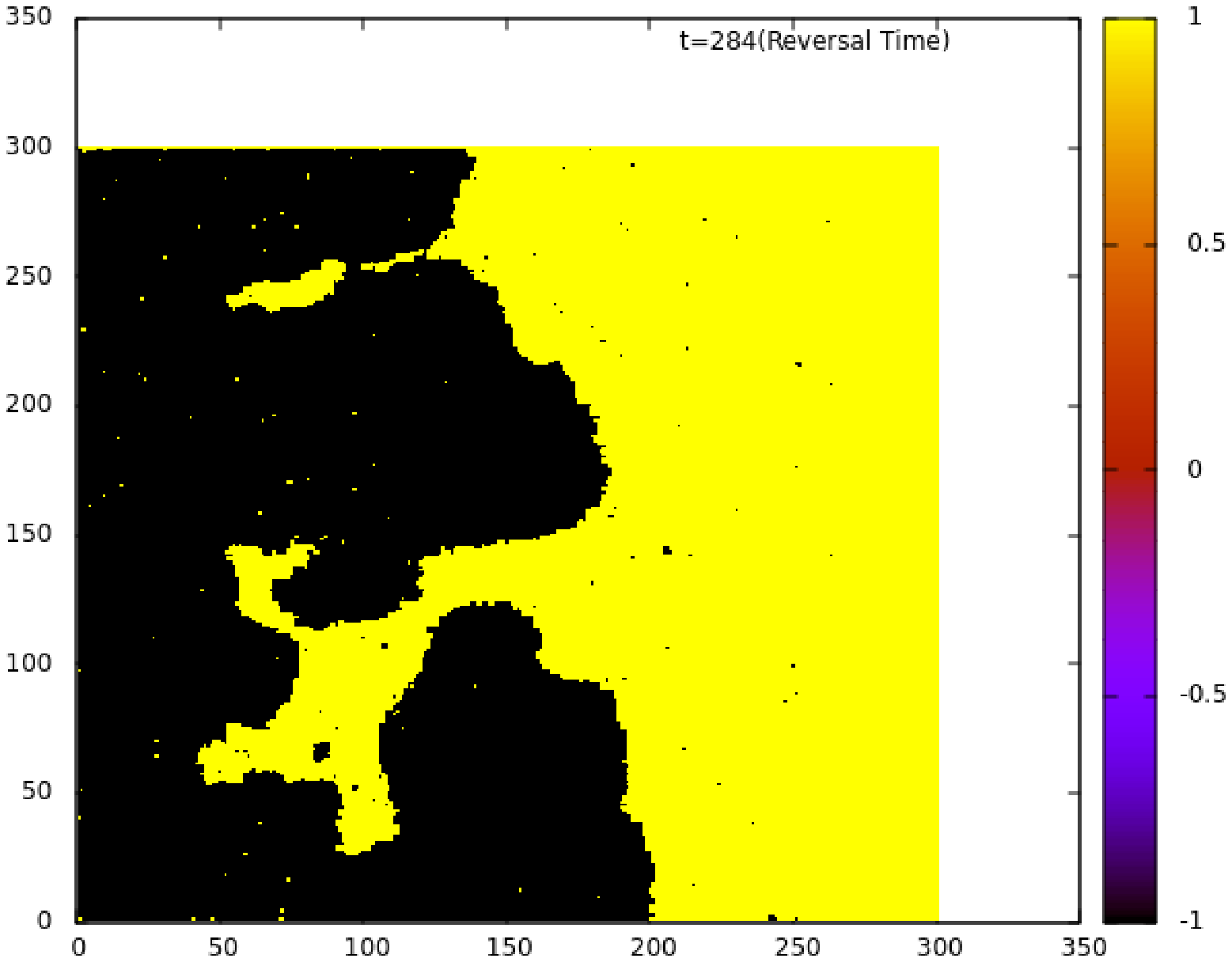}}
d
\resizebox{7cm}{6cm}{\includegraphics[angle=0]{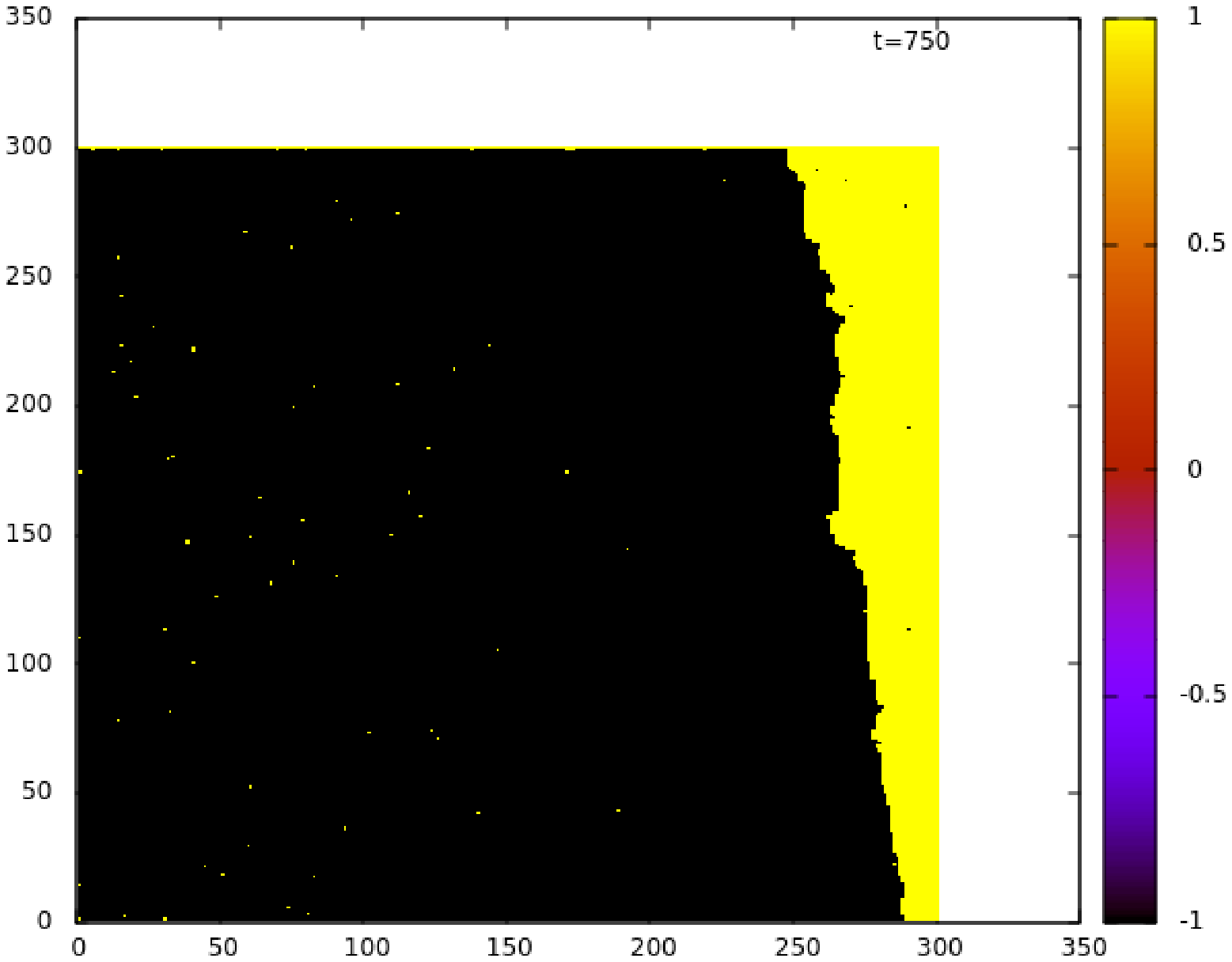}}
          \end{tabular}
\caption{
The evolution of the lattice morphology with time for a system where $T_{l}$=1.4,$T_{r}$=0.9,$h_{r}$=-0.5,$h_{l}$=-0.4. {\bf Top left} (a) The system at t=100.  {\bf Top right} (b) The system at t=200. {\bf Bottom left} (c) The system at t=284 (Reversal Time) . {\bf Bottom right} (d) The system at t=750. Temperatures are measured in the units of $J/k_{B}$ and fields are measured in units of $J$.} 
\label{fig:morph7}
\end{center}
\end{figure}

\begin{figure}[h]
\begin{center}
\begin{tabular}{c}
a
\resizebox{7cm}{6cm}{\includegraphics[angle=0]{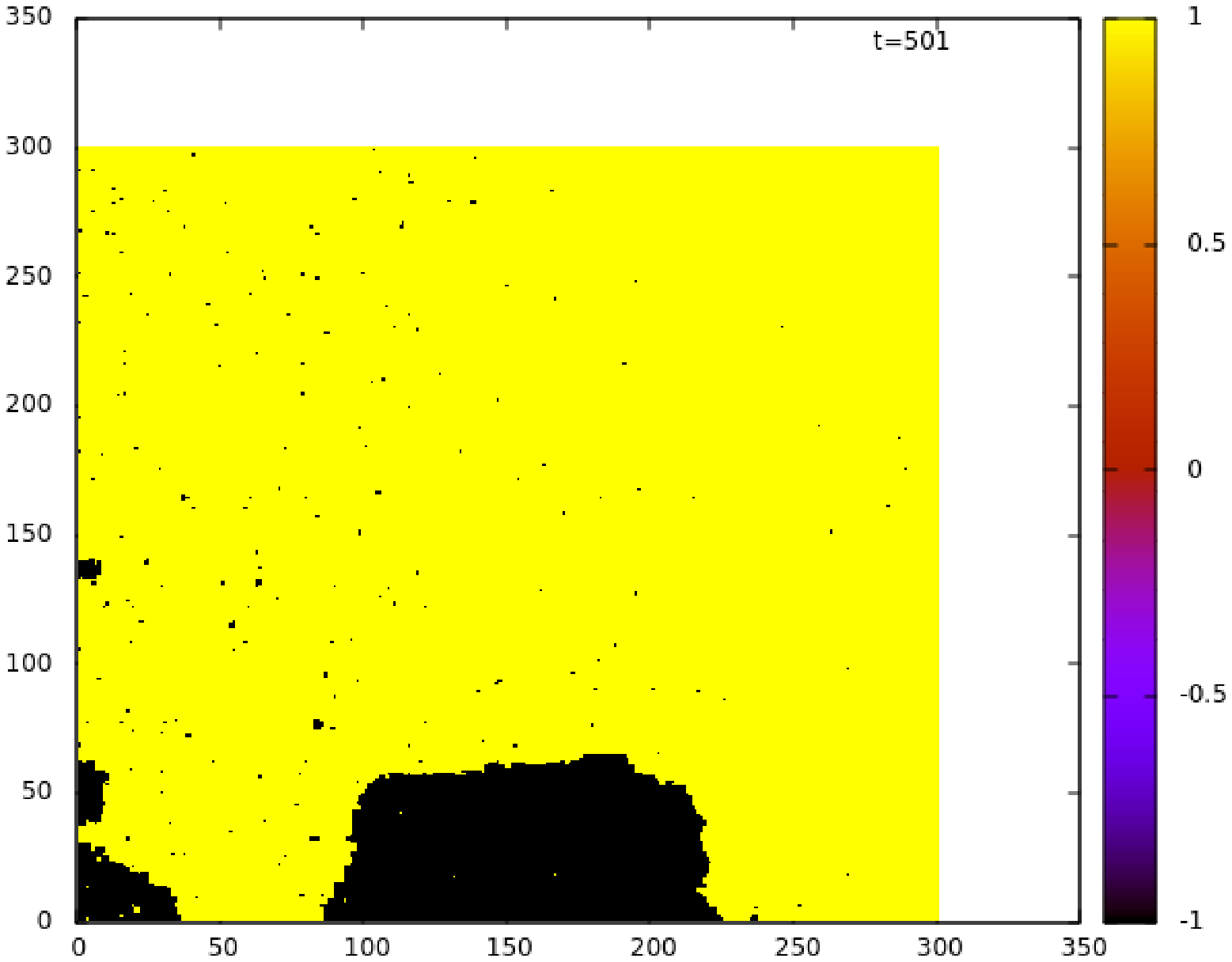}}
b
\resizebox{7cm}{6cm}{\includegraphics[angle=0]{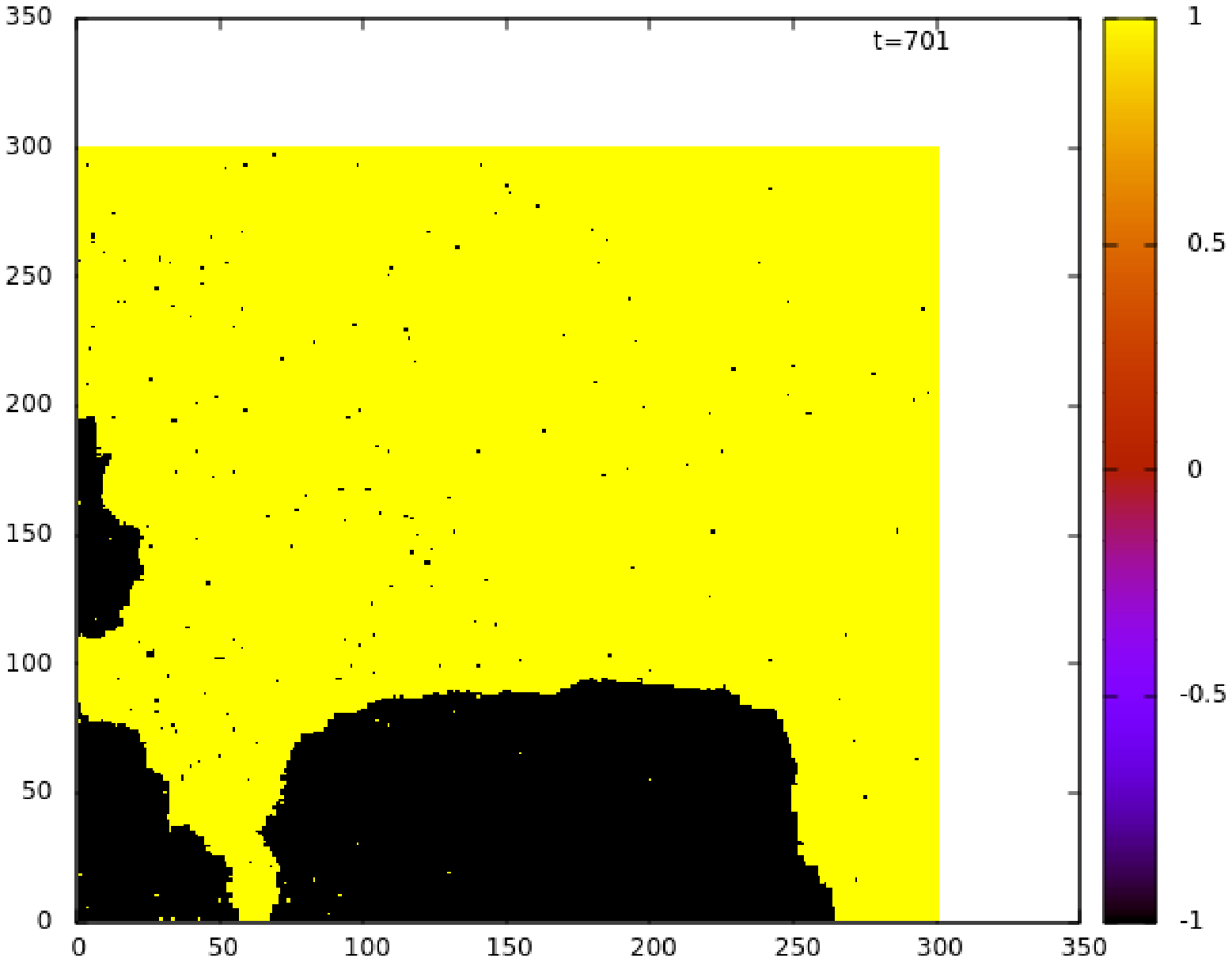}}
\\
c
\resizebox{7cm}{6cm}{\includegraphics[angle=0]{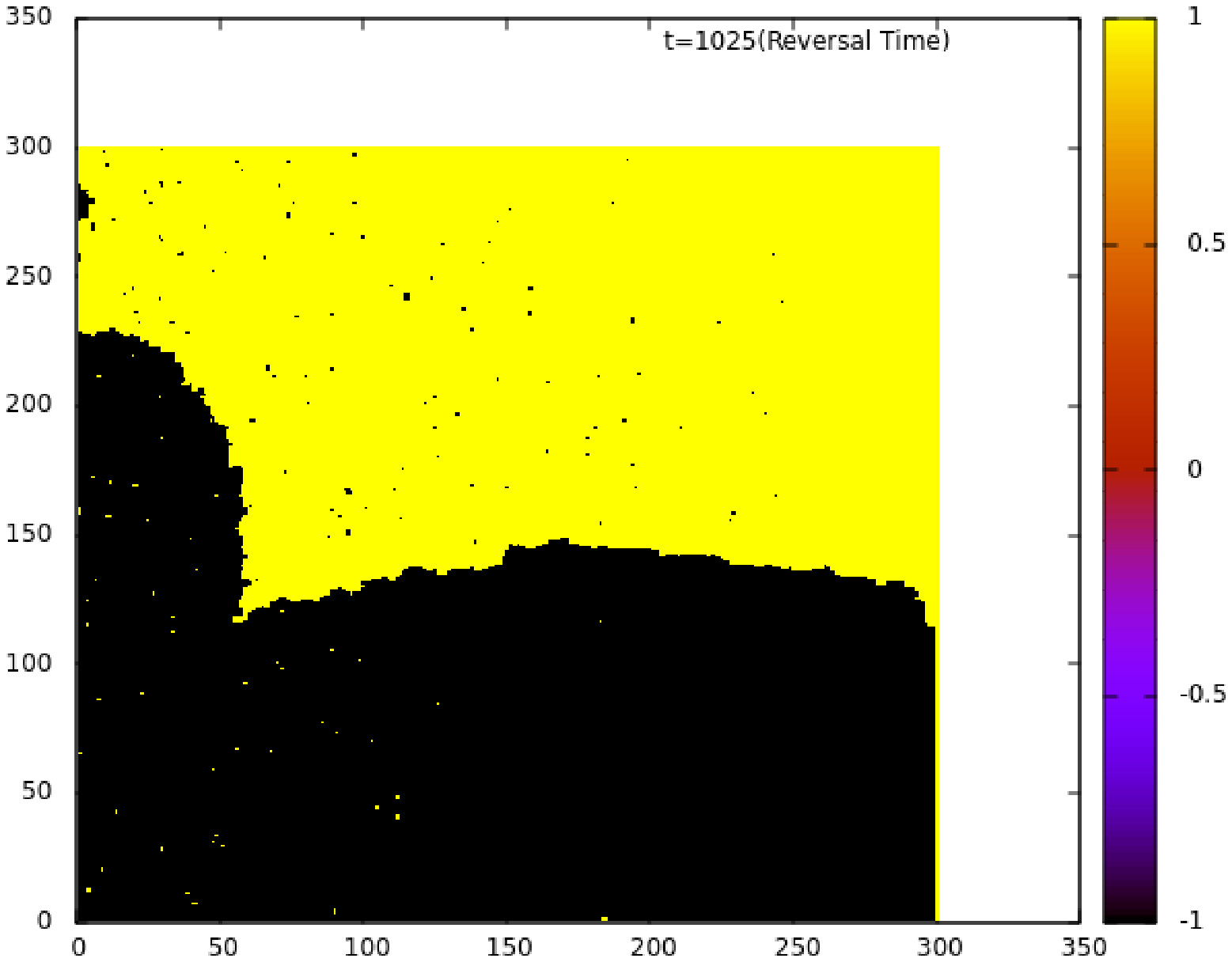}}
d
\resizebox{7cm}{6cm}{\includegraphics[angle=0]{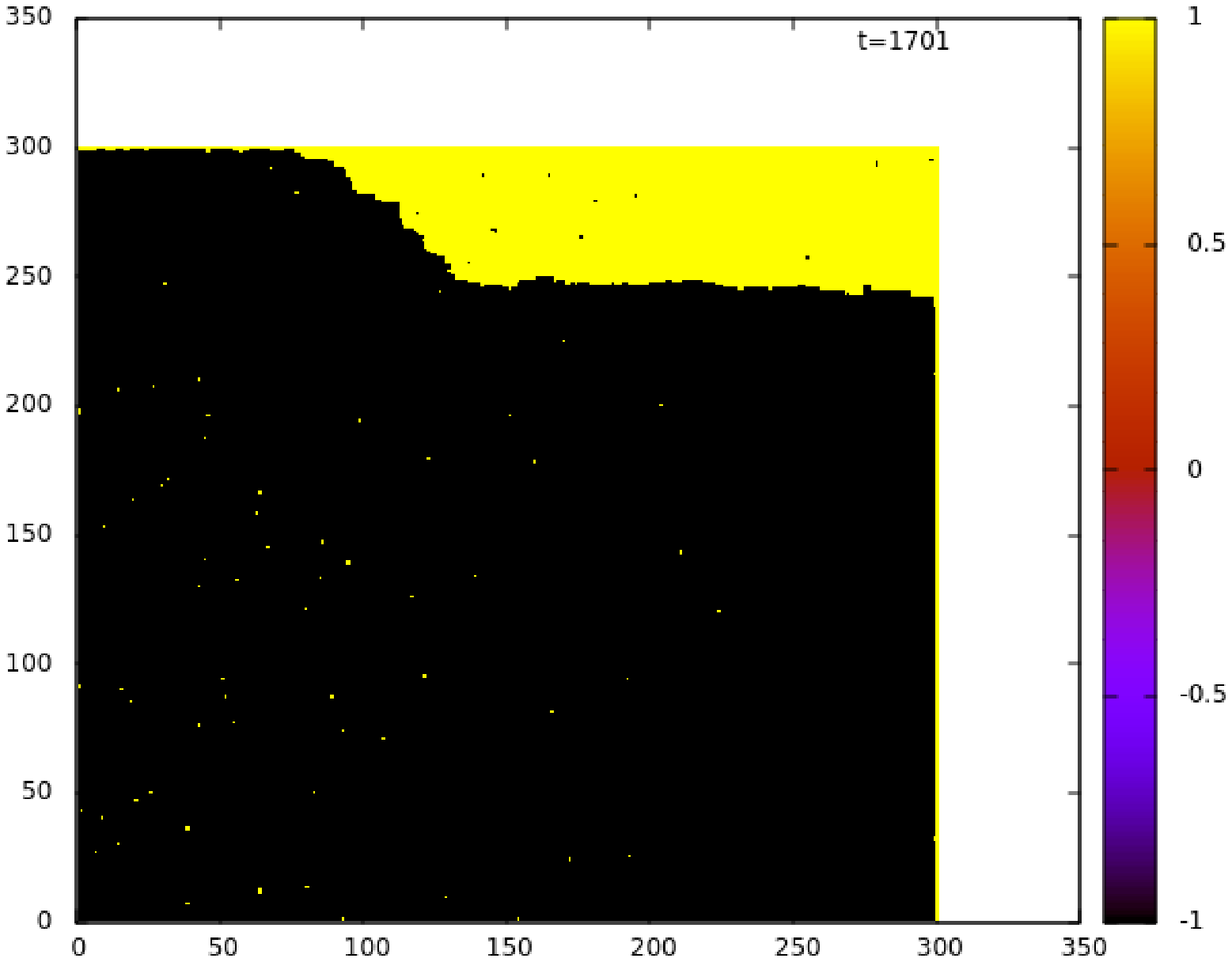}}
          \end{tabular}
\caption{
The evolution of the lattice morphology with time for a system where $T_{l}$=1.4,$T_{r}$=0.9,$h_{r}$=-0.5,$h_{l}$=-0.2. {\bf Top left} (a) The system at t=501.  {\bf Top right} (b) The system at t=701. {\bf Bottom left} (c) The system at t=1025 (Reversal Time) . {\bf Bottom right} (d) The system at t=1701. Temperatures are measured in the units of $J/k_{B}$ and fields are measured in units of $J$.} 
\label{fig:morph8}
\end{center}
\end{figure}

\begin{figure}[h]
\begin{center}
\begin{tabular}{c}
a
\resizebox{7cm}{6cm}{\includegraphics[angle=0]{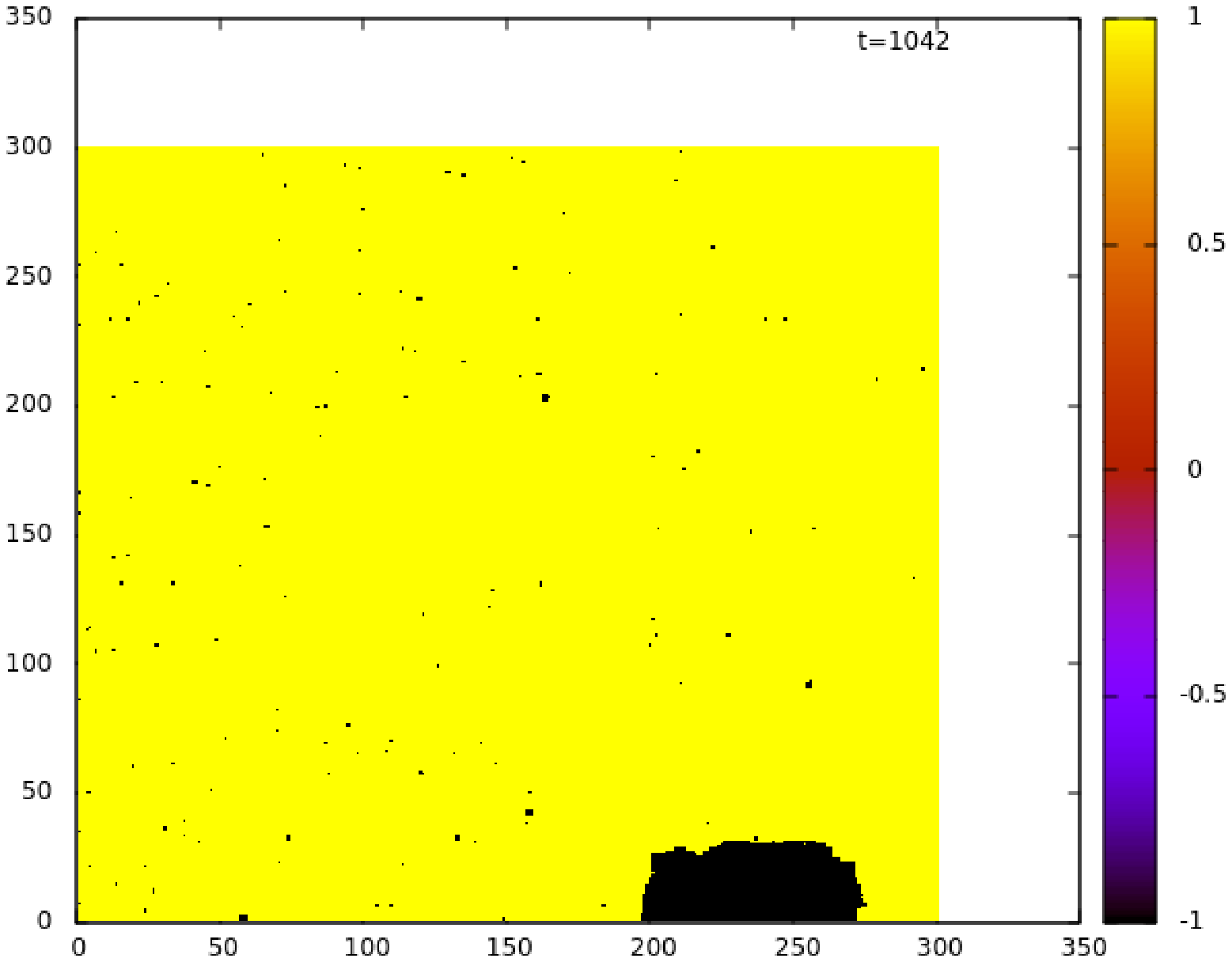}}
b
\resizebox{7cm}{6cm}{\includegraphics[angle=0]{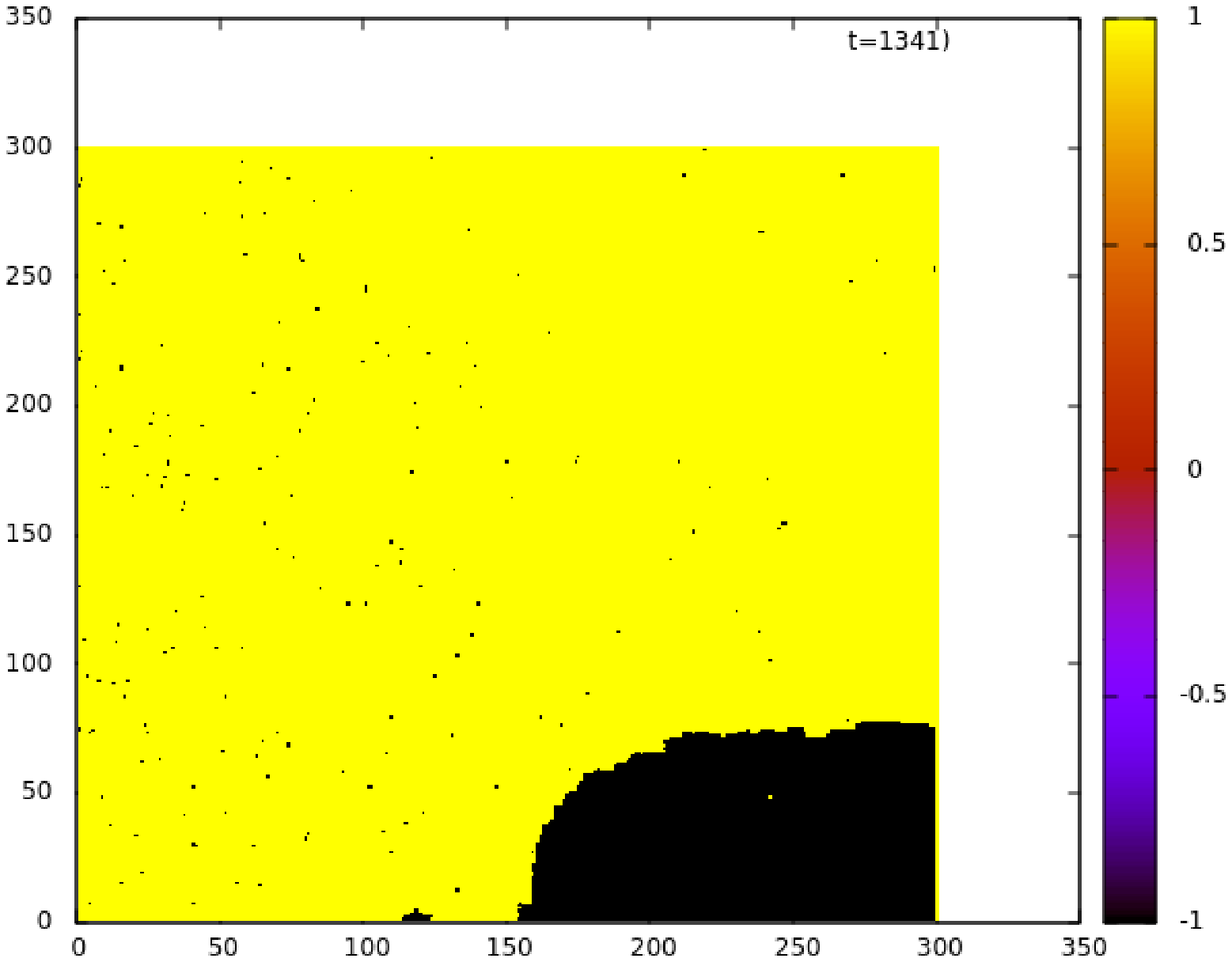}}
\\
c
\resizebox{7cm}{6cm}{\includegraphics[angle=0]{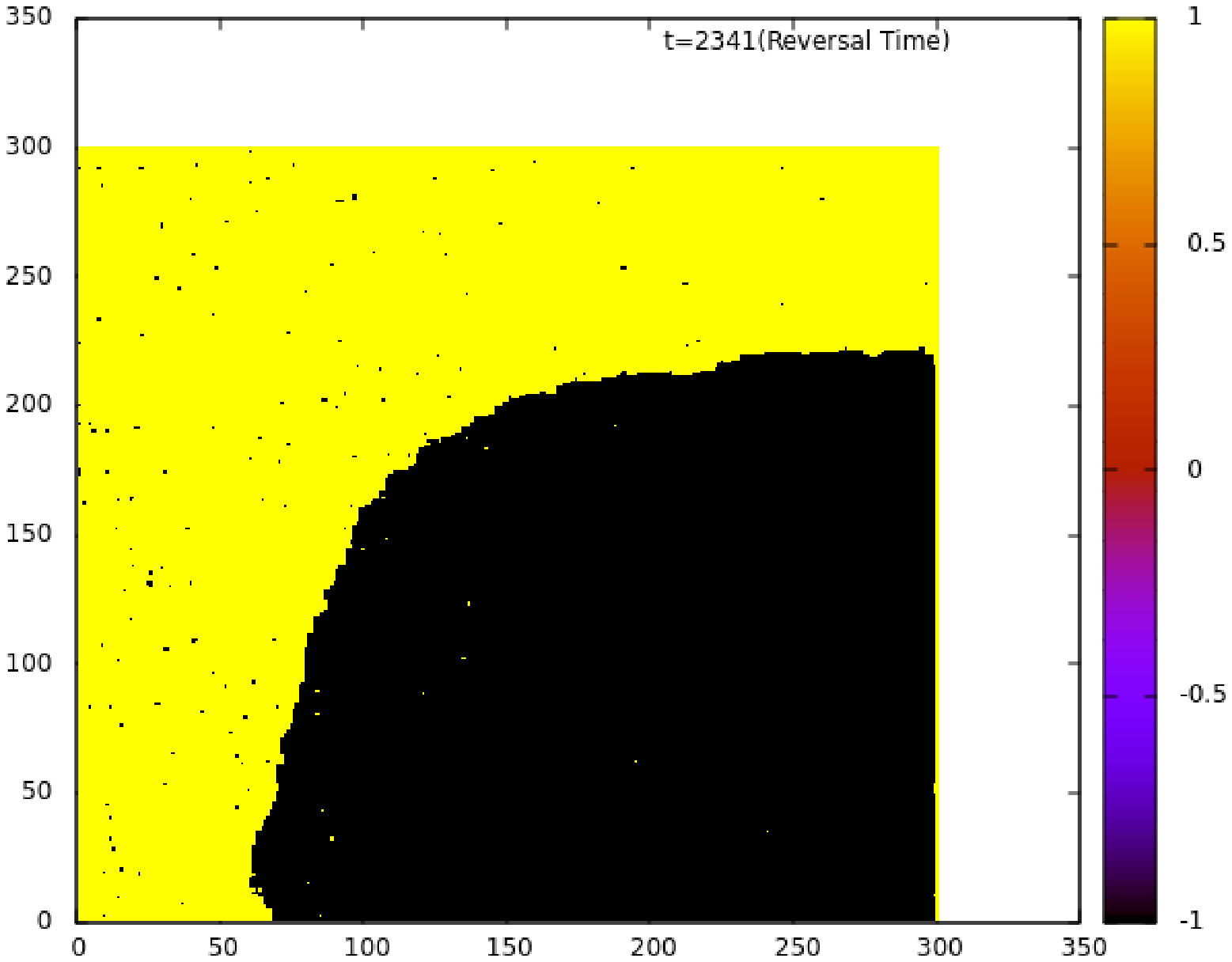}}
d
\resizebox{7cm}{6cm}{\includegraphics[angle=0]{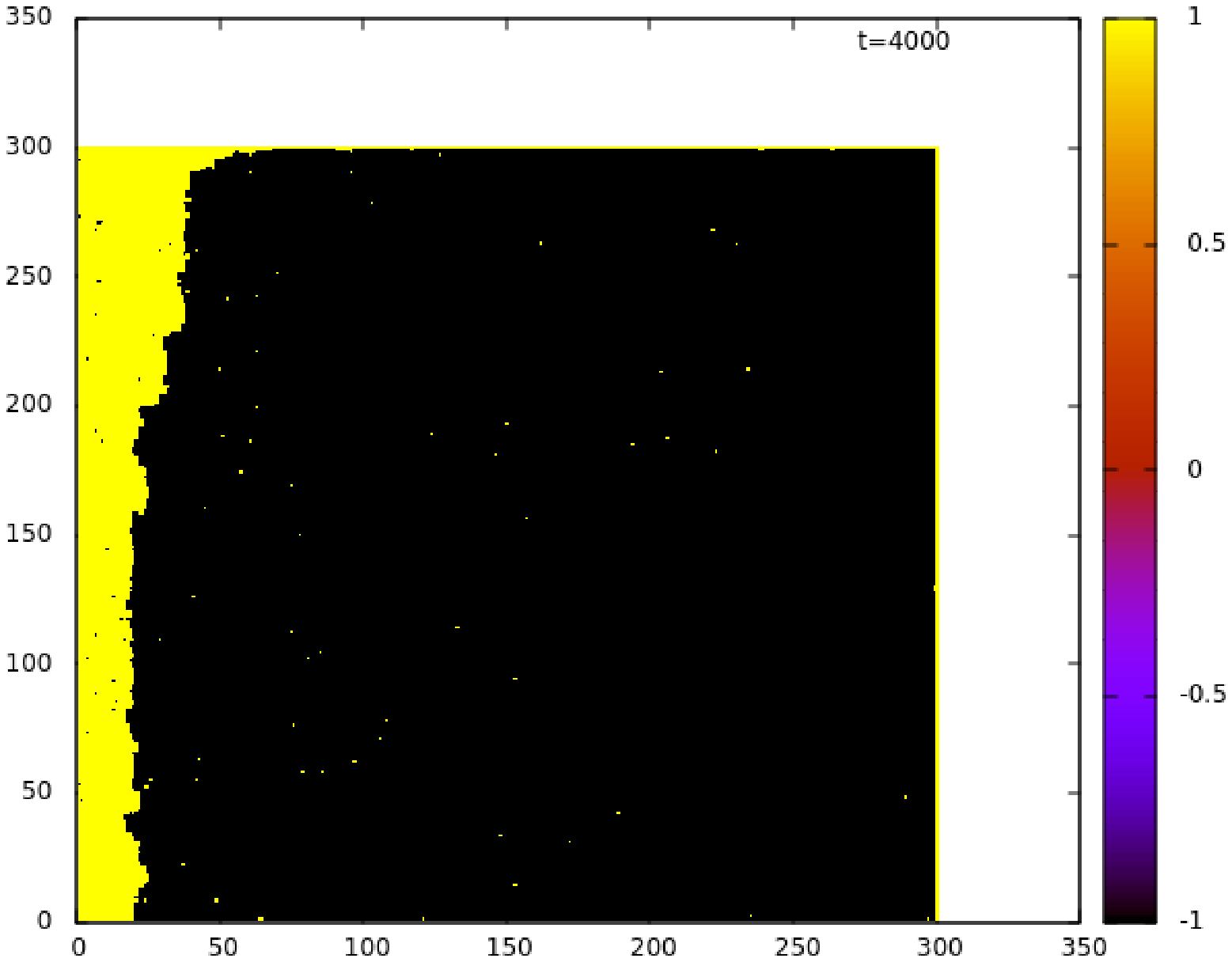}}
          \end{tabular}
\caption{
The evolution of the lattice morphology with time for a system where $T_{l}$=1.4,$T_{r}$=0.9,$h_{r}$=-0.5,$h_{l}$=0.0. {\bf Top left} a) The system at t=1042.  {\bf Top right} b) The system at t=1341. {\bf Bottom left} c) The system at t=2341 (Reversal Time) . {\bf Bottom right} d) The system at t=4000. Temperatures are measured in the units of $J/k_{B}$ and Fields are measured in units of $J$.} 
\label{fig:morph9}
\end{center}
\end{figure}

How does the lattice morphology evolve with time under simultaneous application of thermal and field gradient ? We have studied this and our results are shown in Fig-\ref{fig:morph7}, Fig-\ref{fig:morph8} and Fig-\ref{fig:morph9}.

It is seen that in Fig-\ref{fig:morph7}, where the magnitude of the field at $h_{l}$ is not quite low compared to that of $h_{r}$, multiple clusters (of down spins) are formed and the size of each cluster 
grows as the time progresses. But if we decrease the magnitude of $h_{l}$ compared to that of $h_{r}$, the number of clusters decreases, (as is evident from Fig-\ref{fig:morph8} and Fig-\ref{fig:morph9}.)
And in Fig-\ref{fig:morph9} where $h_{l}$=0.0, only one cluster of negative spin grows. Another way of saying this is as the average magnitude of the negative field decreases, the number of clusters also decreases.

\begin{figure}[h]
\begin{center}
\begin{tabular}{c}
a
\resizebox{7cm}{5cm}{\includegraphics[angle=0]{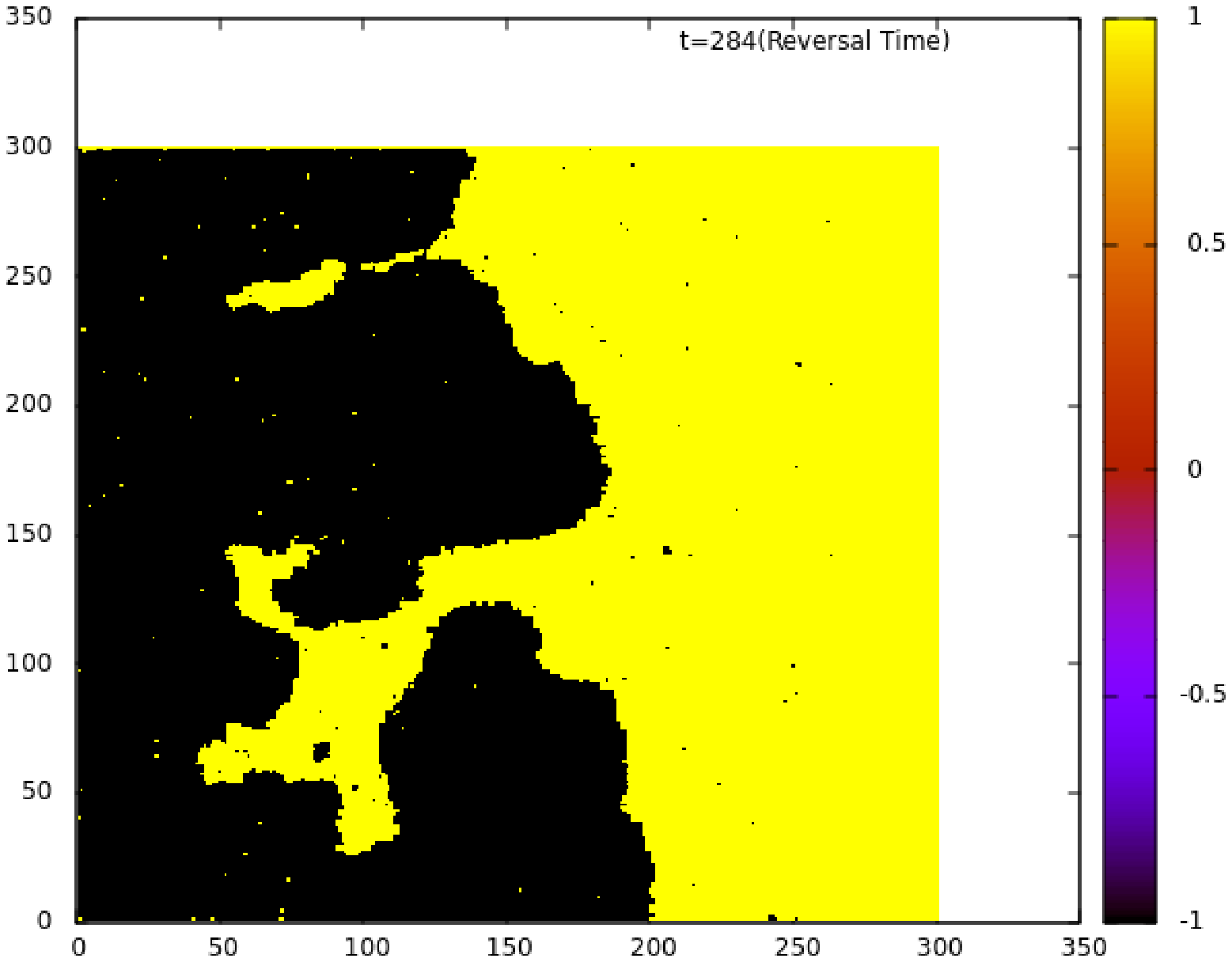}}
\\
b
\resizebox{7cm}{5cm}{\includegraphics[angle=0]{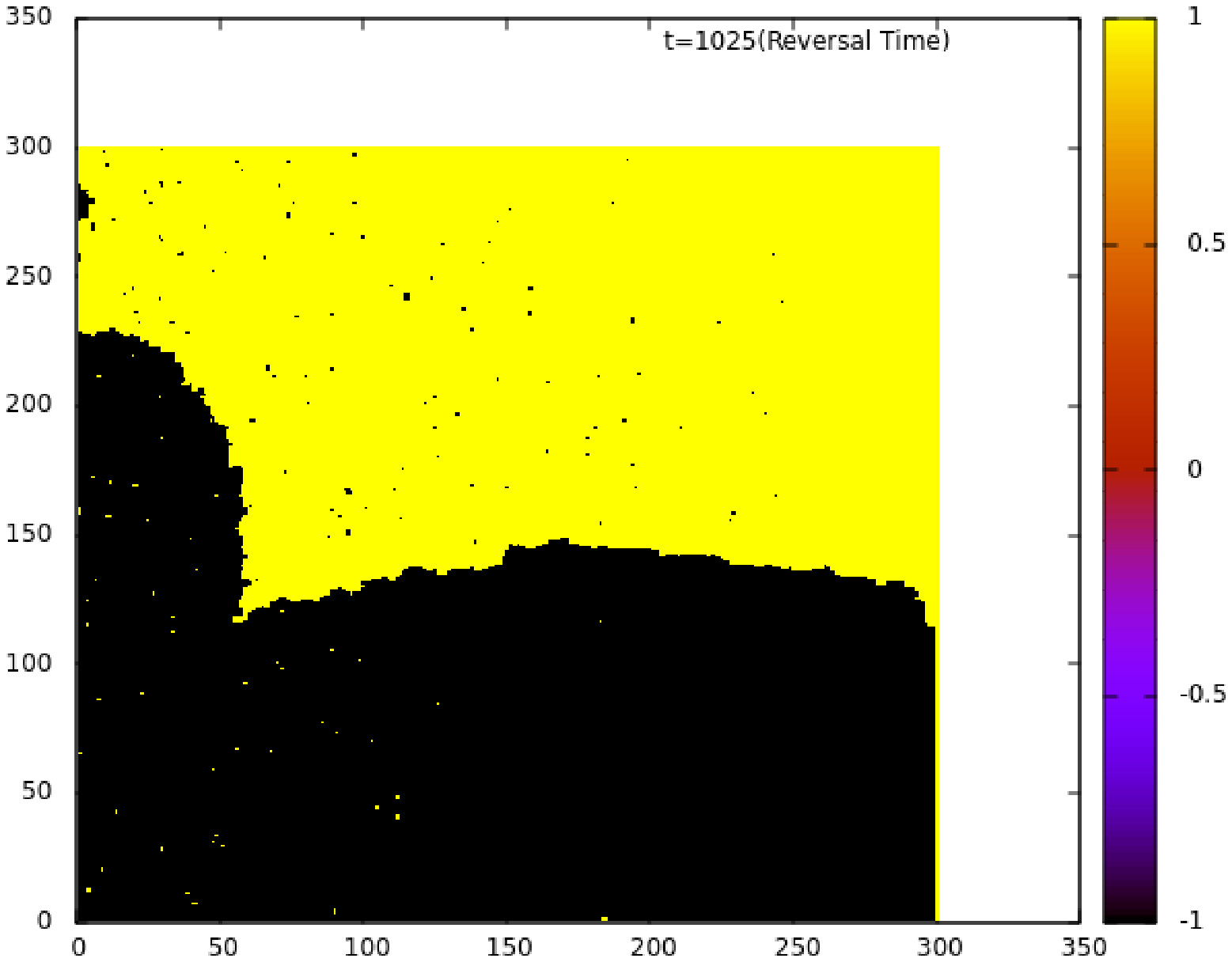}}
\\
c
\resizebox{7cm}{5cm}{\includegraphics[angle=0]{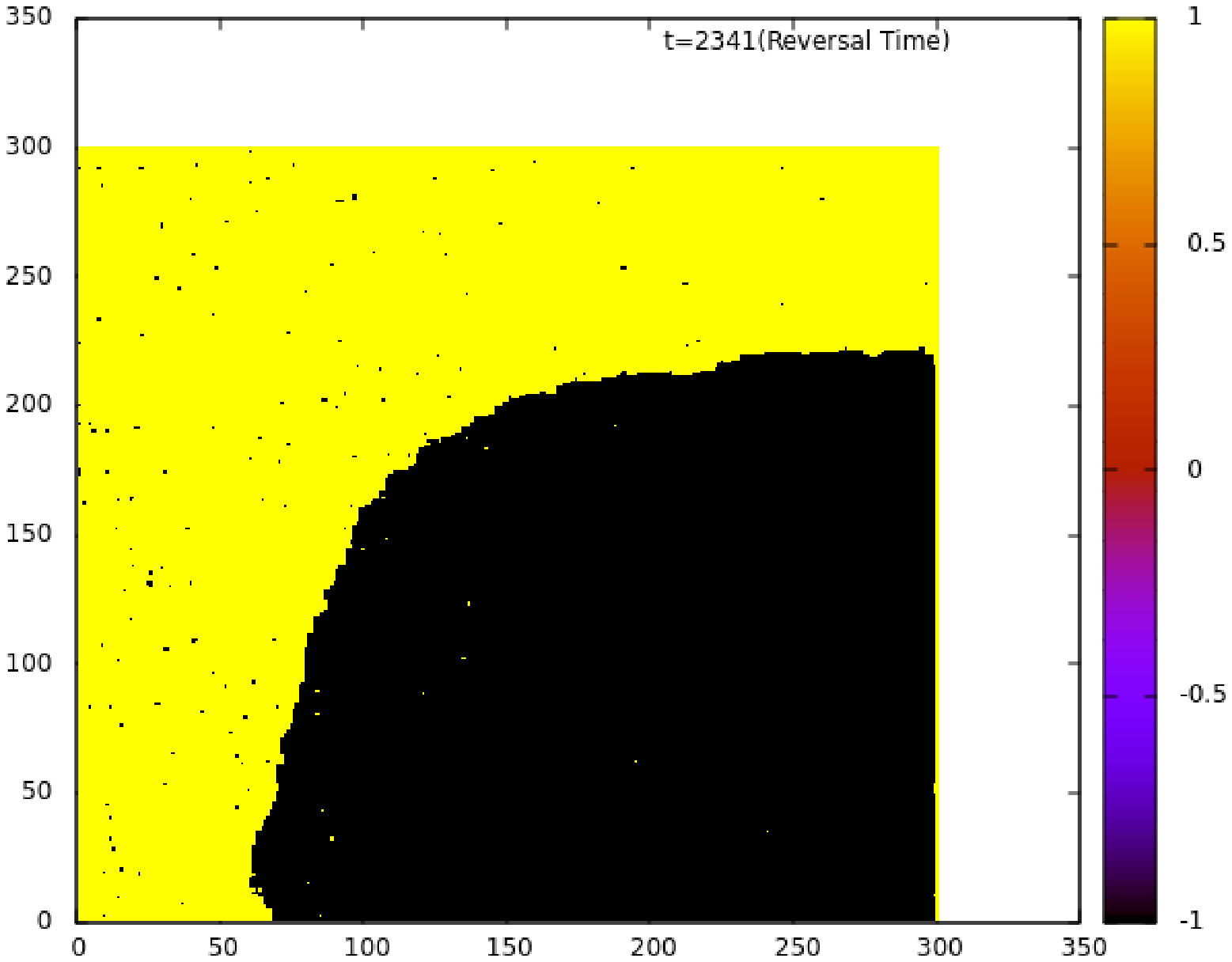}}

          \end{tabular}
\caption{The lattice morphologies (at the time of revbersal) for different sets of thermal gradient and field gradient. {\bf Starting from the top} (a) The system with $T_{l}$=1.4,$T_{r}$=0.9,$h_{r}$=-0.5,$h_{l}$=-0.4. (b) The system with $T_{l}$=1.4,$T_{r}$=0.9,$h_{r}$=-0.5,$h_{l}$=-0.2. (c) The system with $T_{l}$=1.4,$T_{r}$=0.9,$h_{r}$=-0.5,$h_{l}$=0.0.Temperatures are measured in the units of $J/k_{B}$ and fields are measured in units of $J$. } 
\label{fig:morphrev}
\end{center}
\end{figure}

The lattice morphologies at reversal time for these three set of values of temperature and field gradients are compared and shown in Fig-\ref{fig:morphrev}. It is observed that in Fig-\ref{fig:morphrev}a, where the field gradient is small and temperature gradient is comparatively large, at reversal time, all the down spins have clustered near the left edge, signifying that the temperature gradient has dominated over the field gradient. But in Fig-\ref{fig:morphrev}c where the field gradient is high compared to the temperature gradient, at reversal time, all the down spins have clustered near the right edge, signifying the domination of the field gradient over the 
thermal gradient. However in Fig-\ref{fig:morphrev}b, an intermediate phenomenon happens where at reversal time, the clustering of down spins does not show notable affinity towards either the right or the left edge. Here both the gradients are competing equally. We may call it marginal competition between field gradient and the thermal gradient.
Thus the morphology of the lattice at reversal time becomes helpful in indicating which of the gradients dominates over the other. To describe this phenomenon in a quantitative manner, we define a function called the {\bf Competition Factor($C_F$)}. 

\begin{figure}
  \centering
a
    \includegraphics[scale=0.35]{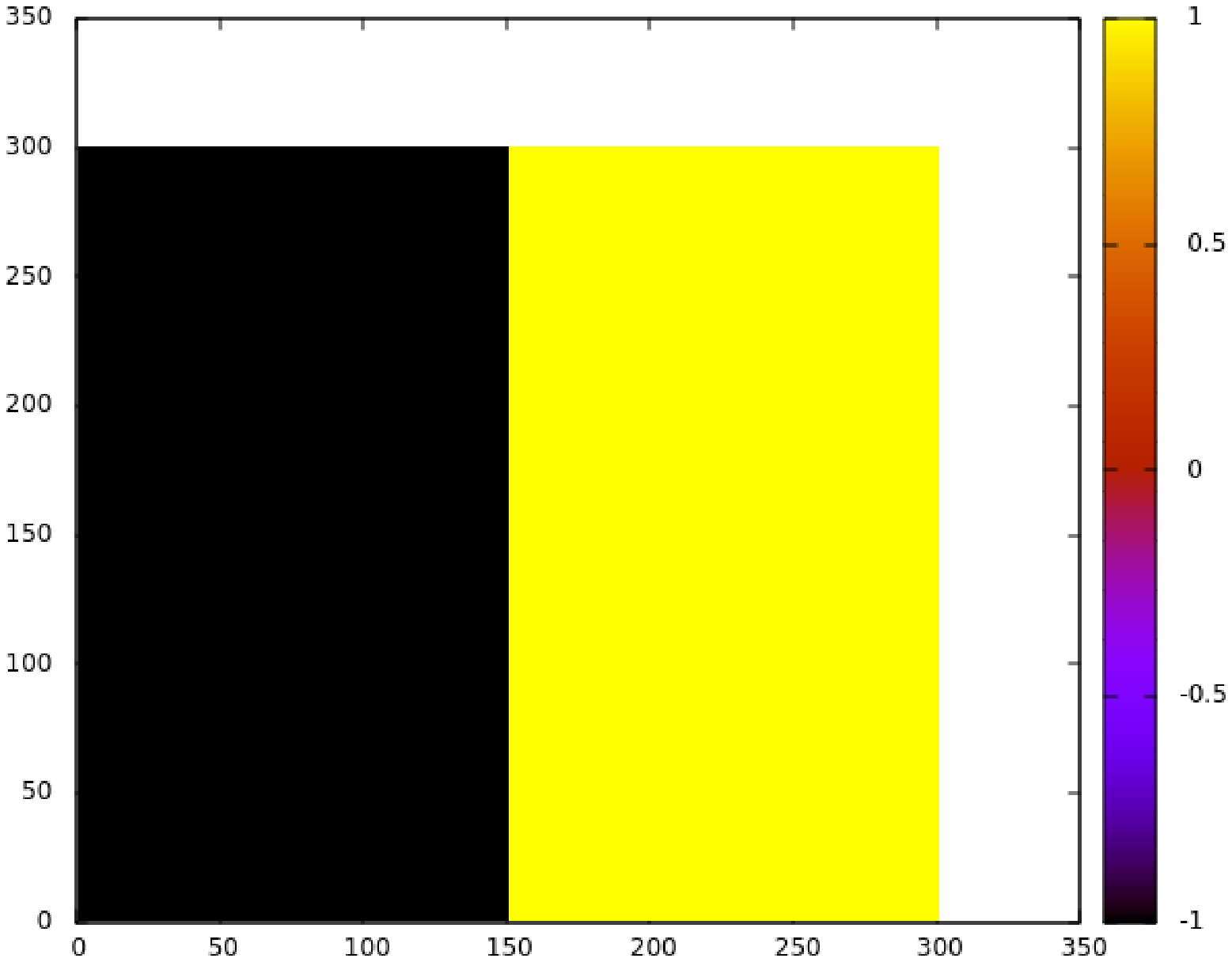}
\\
b
    \includegraphics[scale=0.35]{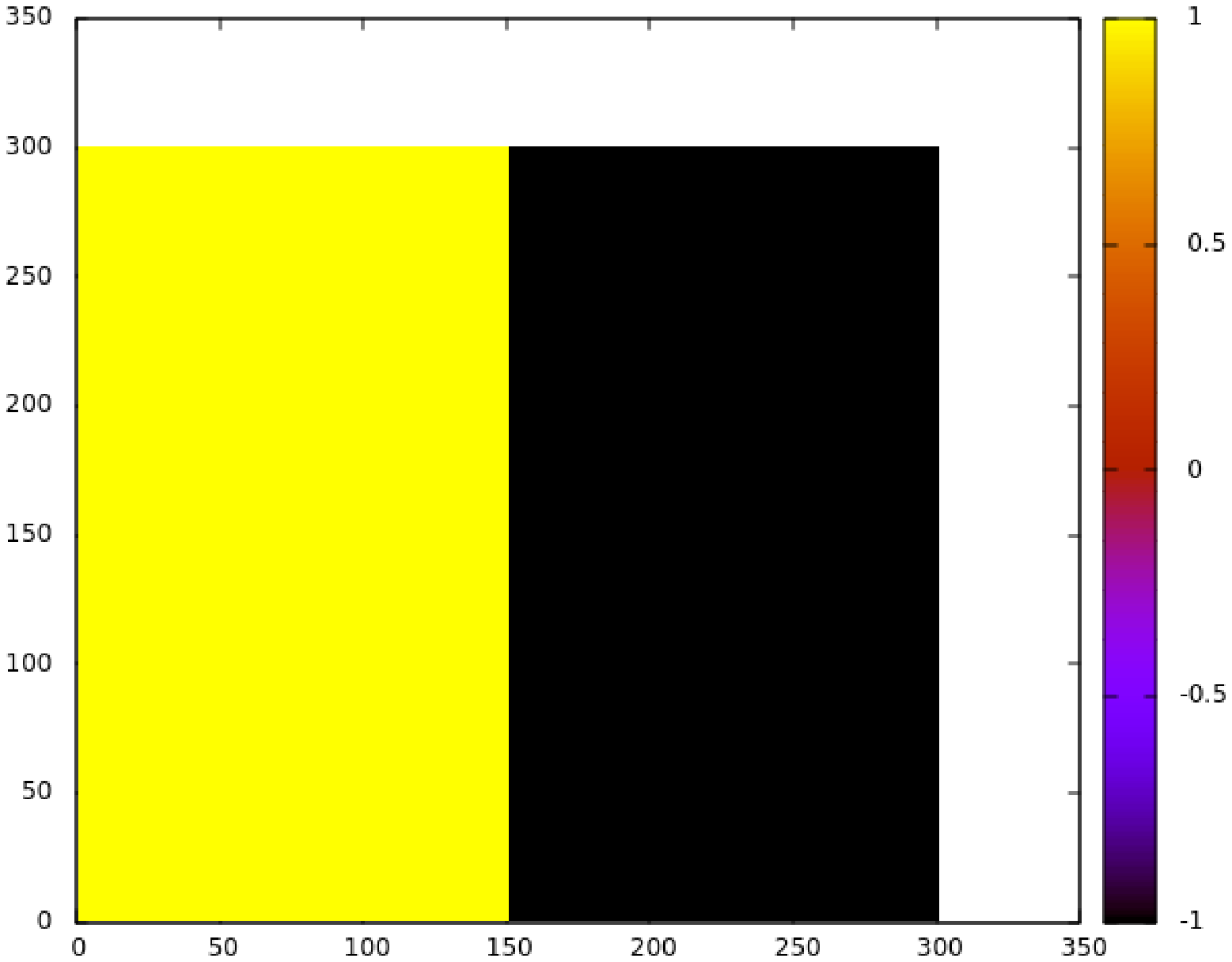}
    
\caption{ {\bf Starting from the top} (a) The reference lattice 
morphology used in the calculation of the $C_F$. This lattice morphology is the manifestation of the ideal case where thermal gradient completely dominates over the field gardient (b) This lattice 
morphology is the manifestation of the ideal case where field gradient completely dominates over the thermal gradient.  }     
\label{fig:refs}
\end{figure}

The morphology in Fig-\ref{fig:refs}a
 can be taken as a manifestation of complete domination of temperature gradient over field gradient. We take this as our reference lattice. The value of the spin  at any site (i,j) on this reference lattice may be denoted by $C(i,j)$. Let $S(i,j)$ represent a spin of any  lattice at reversal time. Then the Competition Factor of this lattice is defined as 
\begin{equation}
    C_F=(\sum_{<i,j>}{C(i,j)*S(i,j)}+(L*L))/(2*L*L)
\end{equation}

The morphology shown in Fig-\ref{fig:refs}a will give a $C_F$ 
(equation-4) of 1. So it is clear that when Temperature Gradient dominates over field gradient, $C_F$ value will be close to 1. Also the morphology shown in Fig-\ref{fig:refs}b, which shows complete domination of field gradient over the thermal gradient, will give a $C_F$ (equation-4) value of 0. So if a system is dominated by field gradient rather than thermal gradient, the $C_F$ value for the morphology is close to 0. However, if for any system,at reversal time, the lattice morphology gives a $C_F$ close to 0.5, that means for that system, both the gradients are competing almost equally (or marginal
competition).

This is clearly demonstrated if we keep the thermal gradient constant and change the field gradient gradually from a lower value to higher values. Table-1 shows when the field gradient is small, the thermal gradient dominates and the $C_F$ is close to 1. And when the field gradient rises, the $C_F$ rises, and when the field gradient becomes so high that it dominates over the temperature gradient, the $C_F$ takes values closer to 0. Also in between these two cases, when the influence of both the gradients are almost equal
(marginal competion), the $C_F$ is close to 0.5.

\begin{figure}[h]

\begin{center}
\begin{tabular}{c}

\resizebox{11cm}{8cm}{\includegraphics[angle=0]{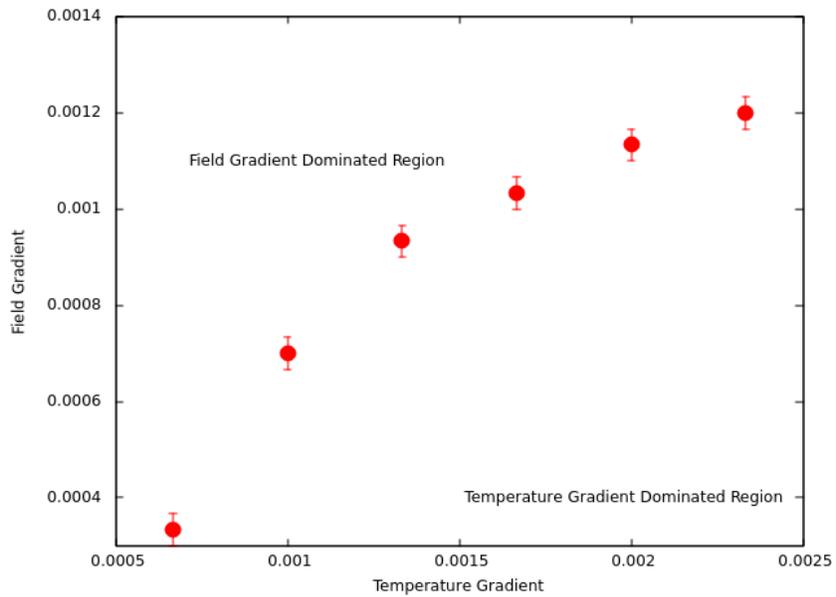}}\\
   \end{tabular} 
    	\caption{ {\bf Line of Marginal Competition:}The line which separates the Field gradient Dominated region from the Temperature Gradient dominated region. Here the y axis has an error bar of 0.0003. Temperatures are measured in the units of $J/k_{B}$ and Fields are measured in units of $J$.} 
\label{fig:mcomp}
\end{center}
\end{figure}

So, keeping the temperature gradient constant, if we vary the field gradient, for every value of the temperature gradient, we will find a value of the field gradient (upto an error bar of 0.0003) for which the $C_F$ is 0.5 (upto an error bar of 0.01). Repeating this process several times, we find a set of thermal gradient and field gradient pairs for which the $C_F$ is very close to 0.5, i.e both the gradients are competing equally (marginal competition). We call the plot of these points the line of marginal Competition (Fig-\ref{fig:mcomp}).This line separates the region of thermal gradient domination from the region of field gradient domination.  

\begin{figure}
  \centering

    \includegraphics[scale=0.6]{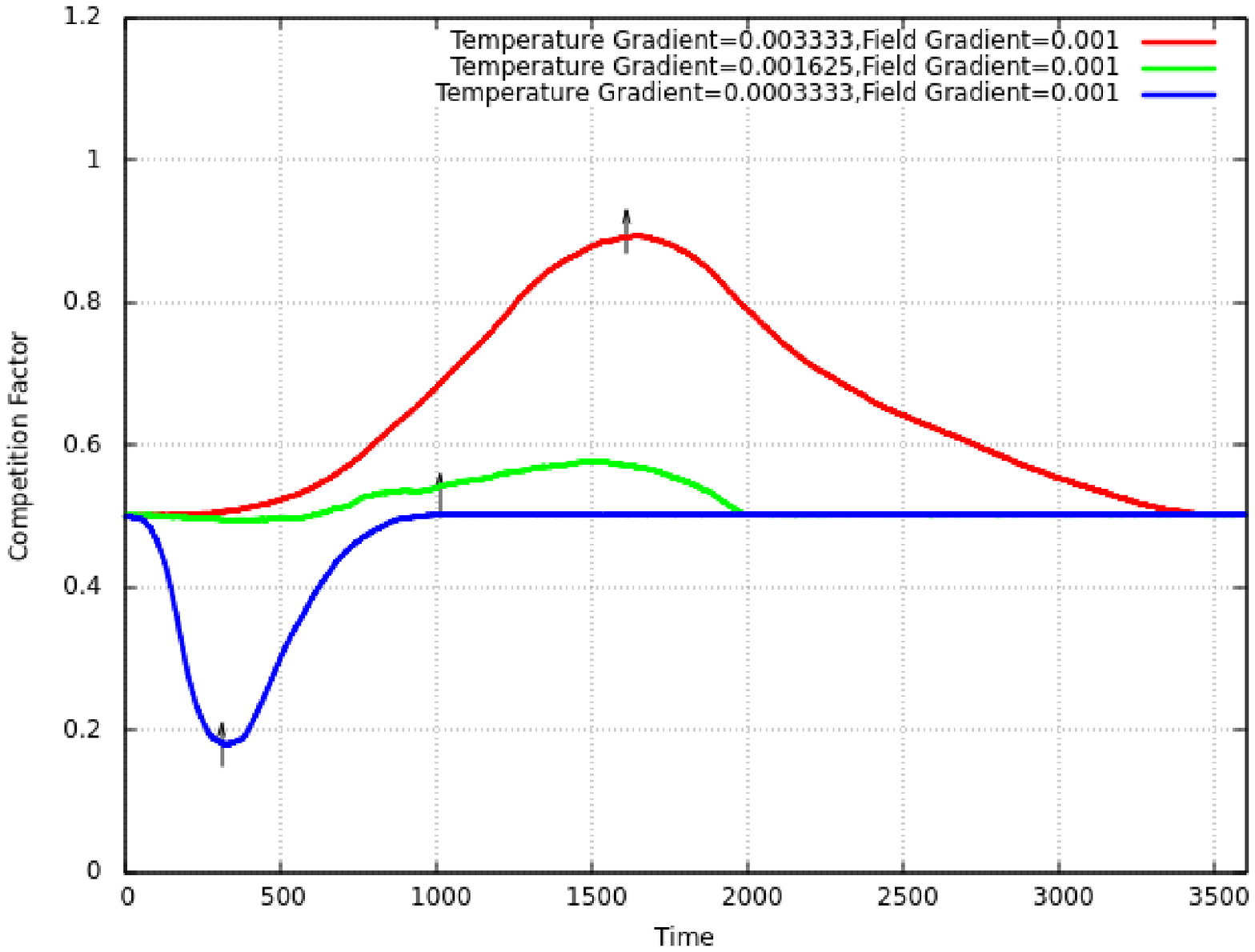}
    
  \caption{ Variation of $C_F$ versus time under 3 different sets of Temperature and Field Gradients(Temperature Gradient Dominated, Field Gradient Dominated,and Intermediate).The arrows mark Reversal Times. Temperatures are measured in the units of $J/k_{B}$ and Fields are measured in units of $J$.} 	  
\label{fig:cf}
\end{figure}

It is worth noting at this point that the Competition Factor is found to be dependent on the lattice morphology at reversal time. It is a worthwhile indicator of which gradient dominates over the other only when the morphology at reversal time is used to calculate the $C_F$. For every system, under any set of conditions, the $C_F$ is 0.5 at the start when all the spins are up, and a long time after reversal when all the spins have flipped, the $C_F$ is again 0.5. Fig-\ref{fig:cf} shows how the $C_F$ varies with time under 3 different sets of conditions.) 

\section{Discussions}

In the first part of this paper, we have investigated the magnetisation 
reversal (by a uniform negative field) and the evolution of the lattice morphology of a two dimensional ising ferromagnet under the influence of a thermal gradient. For our work we had fixed the temperature of one edge of the lattice while varying the temperature of the other edge at values less than the fixed one. We have observed that the average reversal time increases exponentially with increasing gradient. Also, as the reversal time increases, the variance in the reversal time increases. Then we proceeded to study the lattice morphologies of the system at the time 
of reversal, for different values of the thermal gradient. It has been found that for the high value of the thermal gradient, the lattice morphology shows two 
distinct domains of up and down spins, separated by a domain wall. This domain wall becomes smoother as the thermal gradient increases.

In the second part, on top of the thermal gradient, already present, we simultaneously introduce a field gradient in the system. The edge of the lattice having higher temperature is assigned a negative field of lower magnitude. On the other hand, the other edge of the lattice having lower temperature is assigned a negative field of higher magnitude. Thus we created a situation in which both the gradients were made to compete with each other in their spin flipping capability.
Looking at the lattice morphologies at reversal time, for different values of the thermal and field gradients, we can identified which of the two gradients dominated over the other. We defined a mathematical function called the Competition Factor, which, taking into account, the lattice morphology at reversal time, gives a dimensionless number lying between 0 and 1, the value of which determines whether the field gradient dominates or the thermal gradient dominates. Using this function, we found out a set of thermal and field gradient duplets for which both the gradients compete with almost equal strength. We made a plot of such points and called it the line of marginal competition. This line separates the region where thermal gradient dominates from the region dominated by the field gradient.
  
An interesting question may arise regarding the effects of boundary conditions on the roughgness of domain
wall. In our study, we have used the open boundary conditions in all directions. The directions of
both the thermal and field gradients, in our study, are taken along the horizontal axis. As a result,
the values of temperature and field are different in the left and right edges of the lattice.  In a particular column of the square lattice considered here, the
temperature and the fields reamin constant. So, it would
not be justified to apply periodic boundary conditions horizontally. However, it is possible to apply
periodic boundary condition along the vertical direction. Further extensive investigations are required
to check whether the application of such type of periodic boundary condition (along the vertical
direction only) affects the results or not.

The time of reversal of magnetisation will increase as the size of the lattice ($L$) increases. That 
has already been reported\cite{mads} in the case of uniform field and temperature. We believe, in this
case the larger system will increase the time of reversal only. However, this will not remarkably change
the qualitative behaviours upon which we have based our inferences. 
  
The magnetisation reversal in ferromagnetic systems has variety of practical uses. This time of magnetisation reversal is an important parameter for the 
technologies of magnetic recording and storage device\cite{6}. Increase in the reversal time increases their longevity.
In the earlier work\cite{abyaya}, it was shown that increasing the field gradient can result in the increase in reversal time. But in practice, setting up  a field gradient in these devices is difficult.
But it would be easier to set up a thermal gradient and tune it to tune the reversal time. Also if we have simultaneous thermal and field gradient, we have an added degree of freedom, a much more versatile tool in our hand to tune the reversal time.

\newpage
\begin{center} {\bf Table-1} \end{center}
\begin{table}[h!]
\centering
\begin{tabular}{||c c c c c||} 
 \hline
 $T_{l}$ & $T_{r}$ & $h_{r}$ & $h_{l}$ & Average $C_F$  \\ [0.5ex] 
 \hline\hline
 1.4 & 0.9 & -0.5 & -0.4 & 0.8313972 \\ 
 1.4 & 0.9 & -0.5 & -0.3 & 0.7300105 \\
 1.4 & 0.9 & -0.5 & -0.2 & 0.5228405 \\
 1.4 & 0.9 & -0.5 & -0.1 & 0.2742300 \\
 1.4 & 0.9 & -0.5 &  0.0 & 0.2373000 \\ [1ex] 
 \hline
\end{tabular}
\caption{Table showing $C_F$ for different values of thermal and 
field gradients. The $C_F$ is averaged over 100 samples.
Temperatures are measured in the units of $J/k_{B}$ and fields are measured in units of $J$.
The values in the third row are close to the line of marginal competition.}
\label{table:1}
\end{table}

\newpage

\newpage
\newpage

\newpage

\newpage
\newpage

\newpage
\newpage
\newpage
\newpage

\newpage

\newpage

\newpage
\begin{center}{\bf References}\end{center}
\begin{enumerate}
\bibitem{rik0} P. A. Rikvold and B. M. Gorman, 
{\it Recent results on the decay of metastable phases},
Annual Reviews of Computational
Physics I, edited by D. Stauffer (World Scientic, Singapore, 1994),
pp. 149-191;

\bibitem{hanna} H. Venkamaki, {\it Classical nucletation theory in 
multicomponent system}, 2006, Springer-Verlag, Berlin

\bibitem{mads} M. Acharyya and D. Stauffer, 
{\it Nucleation and hysteresis in Ising model : Classical theory versus
computer simulation},
Eur. Phys. J. B {\bf 5}
(1998) 571;\\
 DOI:10.1007/s100510050480

\bibitem{rik1}S. Wonczak, R. Strey, D. Stauffer,
{\it Confirmation of classical nucleation theory by Monte Carlo simulation
in the three dimensional Ising model at low temperature},
J. Chem. Phys. {\bf 113}, 1976 (2000);\\
 DOI:10.1063/1.482003

\bibitem{ryu} S. Ryu and W. Cai, 
{\it Validity of classical nucleation theory for Ising models},
Phys. Rev. E, {\bf 81}, 
030601(R) (2010);\\
 DOI:10.1103/PhysRevE.81.030601 

\bibitem{chen} H. Chen and Z. Hou, 
{\it Optimal modularity for nucleation in a network organised 
Ising model},
Phys. Rev. E, {\bf 83}, 
046124 (2011);\\
DOI:10.1103/PhysRevE.83.046124

\bibitem{bosia} C. Bosia, M. Caselle, D. Cora, 
{\it Nucleation dynamics in two dimensional cylindrical Ising models
and chemotaxis},
Phys Rev E {\bf 81} 
021907 (2010);\\
 DOI:10.1103/PhysRevE.81.021907

\bibitem{bkc} A. Misra and B. K. Chakrabarti, 
{\it Nucleation theory and the phase diagram of the magnetisation reversal
transition},
Europhys. Lett. {\bf 52}
(2000) 311;\\
 DOI:10.1209/epl/i2000-00440-4

\bibitem{ma1} M. Acharyya, Physica A, 
{\it Nucleation in Ising ferromagnet by a field spatially spreading
in time}, Physica A
{\bf 403} (2014) 94;\\
 DOI: 10.1016/j.physa.2014.02.020

\bibitem{rikvold} W. R. Deskins, G. Brown, S. H. Thompson and P. A. Rikvold,
{\it Kinetic Monte Carlo simulation of a model for heat assisted magnetisation
reversal in ultrathin films},
Phys. Rev. B {\bf 84} (2011) 094431;\\
 DOI: 10.1103/PhysRevB.84.094431

\bibitem{uli} U. Nowak, J. Heimel, T. Kleinfeld and D. Weller,
{\it Domain dynamics of magnetic films with perpendicular anisotropy}, 
Phys. Rev. B {\bf 56} (1997) 8143
;\\
 DOI: 10.1103/PhysRevB.56.8143

\bibitem{expt1} C. Bunce, J. Wu, G. Ju, B. Lu, 
D. Hinzke, N. Kazantseva, U. Nowak and R. Chantrell,
{\it Laser induced magnetization switching in films with perpendicular
anisotropy: A comparison between measurements and a multi macro spin model},
Phys. Rev. B, {\bf 81}, 174428 (2010);\\
 DOI: 10.1103/PhysRevB.81.174428

\bibitem{abyaya} A. Dhar and M. Acharyya, 
{\it Reversal of magnetisation in Ising ferromagnet by the field having 
gradient},
Commun. Theor. Phys., {\bf 66} (2016)
563;\\
 DOI: 10.1088/0253-6102/66/5/563

\bibitem{6}C. H. Back, D. Weller, J. Heodmann, 
D. Mauri, D. Guarisco, E. L. Garwin and
H. C. Siegmann, 
{\it Magnetization reversal in ultrashort magnetic field pulses},
Phys. Rev. Lett. {\bf 81}, (1998), 3251;\\
 DOI: 10.1103/PhysRevLett.81.3251

\bibitem{binder} K. Binder and D. W. Heermann, {\it Monte Carlo simulation in
statistical physics}, Springer Series in Solid State Sciences, (Springer,
-York, 1997).
\end{enumerate}
\end{document}